\documentclass[aps,prd,twocolumn,superscriptaddress,nofootinbib,floatfix]{revtex4-2}

\usepackage{xcolor}
\usepackage{float}
\usepackage{graphicx}
\usepackage{amsmath,amssymb,amsfonts}
\usepackage{amsthm}
\usepackage{mathrsfs}
\usepackage{textcomp}
\usepackage{booktabs}
\usepackage{algorithm}
\usepackage{algpseudocode}
\usepackage{listings}
\usepackage[acronym]{glossaries}
\usepackage{tikz}
\usetikzlibrary{positioning,calc,matrix}  
\usepackage{quantikz}
\usepackage{makecell}
\usepackage{multirow}
\usepackage{appendix}

\theoremstyle{thmstyleone}%
\theoremstyle{thmstyletwo}%

\theoremstyle{thmstylethree}%
\glsdisablehyper
\makeglossaries

\newacronym{uboone}{MicroBooNE}{Micro Booster Neutrino Experiment}
\newacronym{mboone}{MiniBooNE}{Mini Booster Neutrino Experiment}
\newacronym{sbn}{SBN}{Short Baseline Neutrino}
\newacronym{sbnd}{SBND}{Short Baseline Near Detector}
\newacronym{dune}{DUNE}{Deep Underground Neutrino Experiment}
\newacronym{fnal}{FNAL}{Fermi National Accelerator Laboratory}
\newacronym{lbnf}{LBNF}{Long-Baseline Neutrino Facility}
\newacronym{nd}{ND}{Near Detector}
\newacronym{fd}{FD}{far detector}
\newacronym{surf}{SURF}{Sanford Underground Research Facility}
\newacronym{cern}{CERN}{Conseil Européen pour la Recherche Nucléaire}
\newacronym{lsnd}{LSND}{Liquid Scintillator Neutrino Detector}
\newacronym{sk}{Super-K}{Super-Kamiokande}
\newacronym{sno}{SNO}{Sudbury Neutrino Observatory}
\newacronym{mi}{MI}{Main Injector}
\newacronym{hk}{Hyper-K}{Hyper-Kamiokande}
\newacronym{gallex}{GALLEX}{Gallium Experiment}
\newacronym{sage}{SAGE}{Soviet-American Gallium Experiment}
\newacronym{sbl}{SBL}{Short-Baseline}
\newacronym{jparc}{JPARC}{Japan Proton Accelerator Research Complex}
\newacronym{iwcd}{IWCD}{Intermediate Water Cherenkov Detector}
\newacronym{reno}{RENO}{Reactor Experiment for Neutrino Oscillation}
\newacronym{neos}{NEOS}{Neutrino Experiment for Oscillation at Short Baseline}

\newacronym{lar}{LAr}{Liquid Argon}
\newacronym{tpc}{TPC}{Time Projection Chamber}
\newacronym{lartpc}{LArTPC}{liquid argon time projection chamber}
\newacronym{lcs}{LCS}{Light Collection System}
\newacronym{crt}{CRT}{Cosmic-Ray Tagger}
\newacronym{apa}{APA}{Anode Plane Assembly}
\newacronym{cpa}{CPA}{Cathode Plane Assembly}
\newacronym{pmt}{PMT}{Photomultiplier Tube}
\newacronym{tpb}{TPB}{Tetraphenyl butadiene}
\newacronym{asic}{ASIC}{Application Specific Integrated Circuit}
\newacronym{daq}{DAQ}{Data Acquisition}
\newacronym{adc}{ADC}{Analogue-to-Digital Converter}
\newacronym{fem}{FEM}{Front End Module}
\newacronym{fpga}{FPGA}{Field-Programmable Gate Array}
\newacronym{tb}{TB}{Trigger Board}
\newacronym{sp}{SP}{Single Phase}
\newacronym{dp}{DP}{Dual Phase}
\newacronym{hd}{HD}{Horizontal Drift}
\newacronym{vd}{VD}{Vertical Drift}
\newacronym{ce}{CE}{Cold Electronics}
\newacronym{pds}{PDS}{Photon Detection System}
\newacronym{fc}{FC}{Field Cage}
\newacronym{arapuca}{ARAPUCA}{Argon R\&D Advanced Program at UniCAmp}
\newacronym{sipm}{SiPM}{silicon photomultiplier}

\newacronym{bnb}{BNB}{Booster Neutrino Beam}
\newacronym{numi}{NuMI}{Neutrinos at the Main Injector}

\newacronym{pot}{POT}{Protons-On-Target}
\newacronym{vuv}{VUV}{Vacuum Ultraviolet}
\newacronym{tbp}{TB}{Test Beam}
\newacronym{cla}{CL}{Confidence Level}

\newacronym{sce}{SCE}{Space-Charge Effect}
\newacronym{edm}{EDM}{Event Data Model}
\newacronym{pfo}{PFO}{Particle-Flow Object}

\newacronym{sm}{SM}{Standard Model}
\newacronym{bsm}{BSM}{Beyond the Standard Model}
\newacronym{rh}{RH}{right-handed}
\newacronym{lh}{LH}{left-handed}
\newacronym{cp}{CP}{charge-parity}
\newacronym{ssm}{SSM}{Standard Solar Model}
\newacronym{pmns}{PMNS}{Pontecorvo–Maki–Nakagawa–Sakata}
\newacronym{msw}{MSW}{Mikheyev-Smirnov-Wolfenstein}
\newacronym{no}{NO}{Normal Ordering}
\newacronym{io}{IO}{Inverted Ordering}
\newacronym{nh}{NH}{Normal Hierarchy}
\newacronym{ih}{IH}{Inverted Hierarchy}
\newacronym{cc}{CC}{Charged Current}
\newacronym{nc}{NC}{Neutral Current}
\newacronym{es}{ES}{Elastic Scattering}
\newacronym{qe}{QE}{Quasi-Elastic}
\newacronym{el}{El}{Elastic}
\newacronym[description={One-particle one-hole}]{1p1h}{1p-1h}{one-particle one-hole}
\newacronym[description={Two-particle two-hole}]{2p2h}{2p-2h}{two-particle two-hole}
\newacronym{mec}{MEC}{Meson Exchange Currents}
\newacronym{coh}{Coh}{Coherent}
\newacronym{res}{Res}{Resonant}
\newacronym{dis}{DIS}{Deep Inelastic Scattering}
\newacronym{ia}{IA}{Impulse Approximation}
\newacronym{rfg}{RFG}{Relativistic Fermi-Gas}
\newacronym{lfg}{LFG}{Local Fermi-Gas}
\newacronym{fsi}{FSI}{Final State Interactions}
\newacronym{sint}{SI}{Secondary Interactions}
\newacronym{ckm}{CKM}{Cabibbo-Kobayashi-Maskawa}
\newacronym{gut}{GUTs}{Grand Unified Theories}

\newacronym{mip}{MIP}{Minimum Ionising Particle}
\newacronym{mcs}{MCS}{Multiple Coulomb Scattering}
\newacronym{ly}{LY}{Light-Yield}
\newacronym{pe}{PE}{Photoelectron}

\newacronym{qed}{QED}{Quantum Electrodynamics}
\newacronym{qcd}{QCD}{Quantum Chromodynamics}
\newacronym{hnl}{HNL}{Heavy Neutral Lepton}
\newacronym{qm}{QM}{Quantum Mechanics}
\newacronym{snu}{SNU}{Solar Neutrino Unit}
\newacronym{si}{SI}{Système International}

\newacronym[description={Muon-neutrino charged-current single charged pion}]{cc1pi}{CC1$\pi^{\pm}$}{muon-neutrino charged-current single charged pion}
\newacronym{fv}{FV}{Fiducial Volume}
\newacronym{pdg}{PDG}{Particle Data Group}

\newacronym{fkr}{FKR}{Feynman-Kislinger-Ravndal}
\newacronym[description={Cosmic Ray}]{cr}{CR}{cosmic ray}
\newacronym{fom}{FoM}{Figure of Merit}
\newacronym{bdt}{BDT}{Boosted Decision Tree}
\newacronym{mc}{MC}{Monte-Carlo}
\newacronym{sw}{SW}{Sanford-Wang}
\newacronym{cv}{CV}{Central-Value}
\newacronym{pg}{PG}{Probability-of-Goodness}
\newacronym{red}{R$\&$D}{Research and Development}
\newacronym{ad}{AD}{Antineutrino Detector}
\newacronym{iw}{IW}{Inner Water}
\newacronym{ow}{OW}{Outer Water}
\newacronym{rpc}{RPC}{Resistive Plate Chamber}

\newacronym{ai}{AI}{Artificial Intelligence}
\newacronym{ml}{ML}{machine learning}
\newacronym{dl}{DL}{deep learning}
\newacronym{qml}{QML}{quantum machine learning}
\newacronym{ann}{ANN}{Artificial Neural Network}
\newacronym{mlp}{MLP}{multi-layer perceptron}
\newacronym{cnn}{CNN}{Convolutional Neural Network}
\newacronym{nn}{NN}{}
\newacronym{pqc}{PQC}{parametrised quantum circuit}
\newacronym{qcnn}{QCNN}{quantum convolutional neural network}
\newacronym{cpu}{CPU}{Central Processing Unit}
\newacronym{gpu}{GPU}{Graphic Processing Unit}
\newacronym{tpu}{TPU}{Tensor Processing Unit}
\newacronym{tf}{Tf}{TensorFlow}
\newacronym{iot}{IoT}{Internet-of-Things}
\newacronym{ram}{RAM}{Random Access Memory}
\newacronym{api}{API}{Application Programming Interface}
\newacronym{qat}{QAT}{Quantisation Aware Training}
\newacronym{tdp}{TDP}{Thermal Design Power}
\newacronym{vram}{VRAM}{Video Random Access Memory}
\newacronym{hz}{Hz}{hertz}
\newacronym{bios}{BIOS}{Basic Input/Output System}
\newacronym{rom}{ROM}{Read-Only Memory}
\newacronym{alu}{ALU}{arithmetic logic unit}
\newacronym{fpu}{FPU}{Floating Point Unit}
\newacronym{mac}{MAC}{Multiply-Accumulate}
\newacronym{mxu}{MXU}{Matrix Multiplication Unit}
\newacronym{hbm}{HBM}{High Bandwidth Memory}
\newacronym{npu}{NPU}{Neural Processing Unit}
\newacronym{mosfet}{MOSFET}{metal–oxide–semiconductor field-effect 
transistors}
\newacronym{tmva}{TMVA}{Toolkit for Multivariate Data Analysis}
\newacronym{mva}{MVA}{Multivariate Data Analysis}
\newacronym{resnet}{ResNet}{Residual Neural Network}
\newacronym{dense}{DenseNet}{Densely Connected Convolutional Network}
\newacronym{ptiq}{PTIQ}{Post-Training Integer Quantisation}
\newacronym{roc}{ROC}{Receiver Operating Characteristic}

\newacronym{pmf}{pmf}{probability mass function}
\newacronym{cdf}{cdf}{cumulative distribution function}
\newacronym{pdf}{pdf}{probability density function}
\newacronym{clt}{CLT}{central limit theorem}
\newacronym{pca}{PCA}{Principal Component Analysis}
\newacronym{svm}{SVM}{Support Vector Machine}
\newacronym{lda}{LDA}{Linear Discriminant Analysis}
\newacronym{mle}{MLE}{Maximum Likelihood Estimation}
\newacronym{sgd}{SGD}{Stochastic Gradient Descent}
\newacronym{dof}{dof}{degrees of freedom}

\newacronym{tki}{TKI}{Transverse Kinematic Imbalance}
\newacronym{tmi}{TMI}{Transverse Momentum Imbalance}
\newacronym{tba}{TBA}{Transverse Boosting Angle}
\newacronym{aba}{ABA}{Azimuthal Boosting Angle}
\newacronym{cl}{CL}{Collaborative Learning}
\newacronym{hep}{HEP}{high energy physics}

\newacronym{mit}{MIT}{Massachusetts Institute of Technology}

\newacronym{su}{SU}{Special Unitary}
\newacronym{ssb}{SSB}{Spontaneous symmetry breaking}

\begin{document}
\twocolumngrid

\title[Article Title]{LArTPC hit-based topology classification with quantum machine learning and symmetry}














\author{Callum Duffy}
\thanks{These authors contributed equally to this work.}
\affiliation{Department of Physics and Astronomy, University College London, London, UK}

\author{Marcin Jastrzebski}
\thanks{These authors contributed equally to this work.}
\affiliation{Department of Physics and Astronomy, University College London, London, UK}

\author{Stefano Vergani}
\email{s.vergani@ucl.ac.uk}
\affiliation{Department of Physics and Astronomy, University College London, London, UK}

\author{Leigh H. Whitehead}
\affiliation{Cavendish Laboratory, University of Cambridge, Cambridge, UK}

\author{Ryan Cross}
\affiliation{Department of Physics, University of Warwick, Coventry, UK}

\author{Andrew Blake}
\affiliation{Physics Department, Lancaster University, Lancaster, UK}

\author{Sarah Malik}
\affiliation{Department of Physics and Astronomy, University College London, London, UK}

\author{John Marshall}
\affiliation{Department of Physics, University of Warwick, Coventry, UK}


\begin{abstract}
We present a new approach to separate track-like and shower-like topologies in \gls{lartpc} experiments for neutrino physics using quantum machine learning. Effective reconstruction of neutrino events in LArTPCs requires accurate and granular information about the energy deposited in the detector. These energy deposits can be viewed as 2-D images. Simulated data from the MicroBooNE experiment and a simple custom dataset are used to perform pixel-level classification of the underlying particle topology. Images of the events have been studied by creating small patches around each pixel to characterise its topology based on its immediate neighbourhood. This classification is achieved using convolution-based learning models, including quantum-enhanced architectures known as quanvolutional neural networks. The quanvolutional networks are extended to symmetries beyond translation. Rotational symmetry has been incorporated into a subset of the models. This study demonstrates that quantum-enhanced models perform better than their classical counterparts with a comparable number of parameters, but are outperformed by classical models with two orders of magnitude more parameters. The inclusion of rotation symmetry appears benefitial only in a small number of cases and remains to be explored further. Possible future use of quantum machine learning in the reconstruction phase is discussed, with emphasis on future \gls{lartpc} experiments such as \gls{dune}-\gls{fd}.

\end{abstract}




\maketitle

\section{Introduction}\label{sec:intro}

\begin{figure}[h!]
    \centering   
    \includegraphics[width = \linewidth]{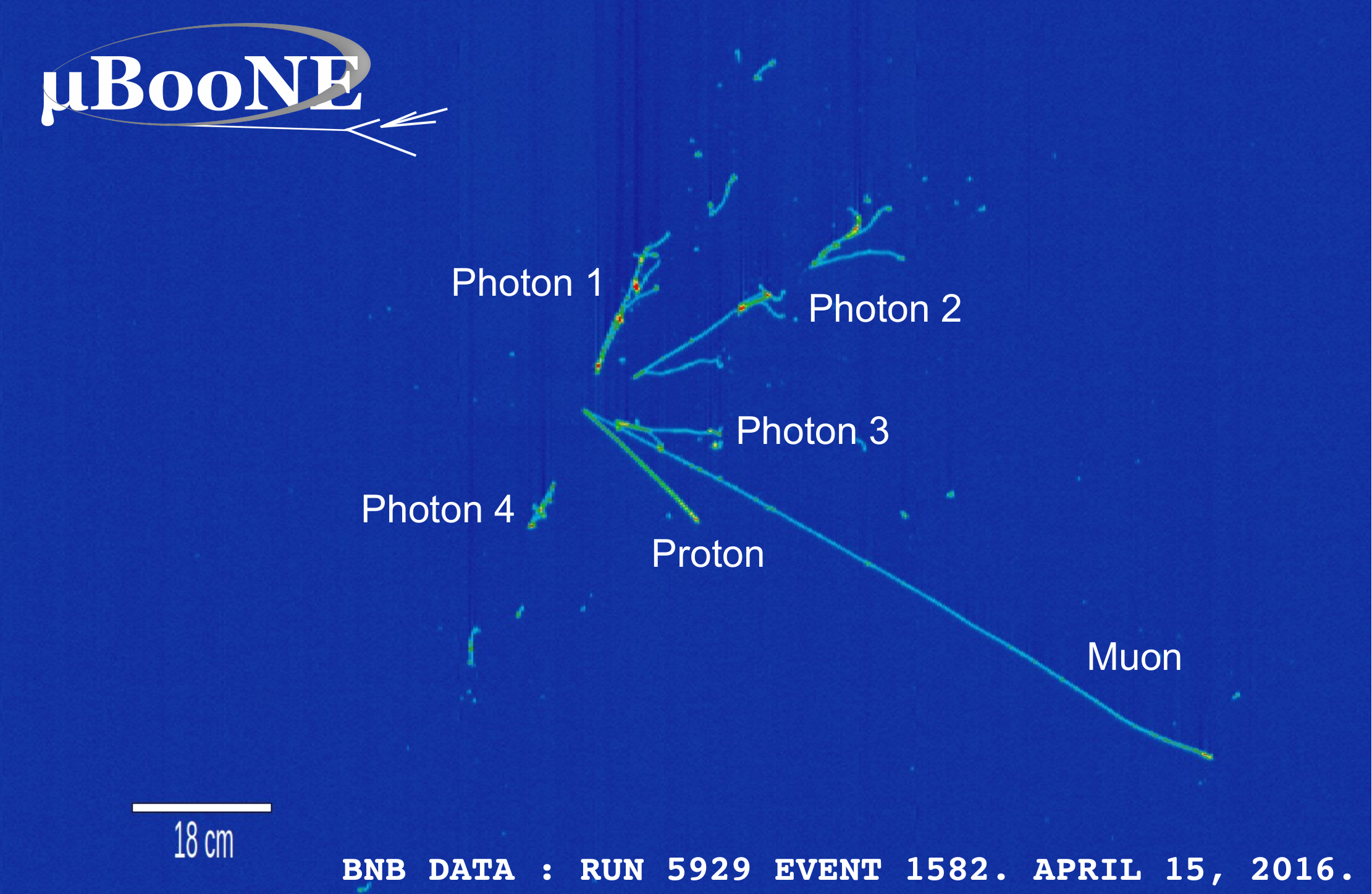}
    \caption{An example neutrino interaction event viewed in a \gls{lartpc} display from the \gls{uboone} experiment. Six particles can be seen emerging from the interaction vertex: two track-like (muon and proton) and four shower-like (photons). Figure adapted from Ref.~\cite{uboone_display}} 
\label{fig:microboone_display}
\end{figure}

\gls{lartpc} detectors are commonly used in neutrino physics and represent the current state of the art of in the field. Although there are differences in the design, they normally comprise a cryostat filled with high-purity liquid argon kept at 87\,K, a system to apply an internal electric field, a readout mechanism made with wires, and a photon detection system. In \gls{lartpc}s, charged particles interact with argon nuclei, liberating electrons through ionisation. These electrons drift in an electric field towards readout wires, where they are detected. Combining the wire number with the time of detection, especially if the scintillation light is used as a trigger, gives spatial and time information in a 2-D view. Calorimetric information is added to each 2-D view. The interactions of neutrinos with the argon nuclei are inferred by the presence of charged particles produced in the interaction emerging from a common vertex. An example of a 2-D image produced by a neutrino interaction in a \gls{lartpc} experiment is shown in Figure \ref{fig:microboone_display}.
The \gls{uboone} \cite{Acciarri_2017} employed an 85 metric tons \gls{lartpc} detector to perform precision physics measurements and collected five years of data from 2016 to 2021. The experiment's primary scientific goals were to solve the puzzle of the MiniBooNE low energy excess \cite{Arg_elles_2022}, to measure different neutrino cross sections, and to search for astrophysical phenomena. It was part of the wider \gls{sbn} \cite{Machado_2019}, an ambitious program at \gls{fnal} aimed at investigating eV-scale sterile neutrinos, neutrino-nucleus interactions at the GeV energy scale, and the advancement of the liquid argon detector technology. The \gls{dune} \cite{DUNE:2020jqi} represents the next generation of neutrino physics experiments, and is partly located at the \gls{lbnf} at \gls{fnal} and partly at \gls{surf}. It aims to answer fundamental questions in particle physics, such as the neutrino mass ordering \cite{Fukasawa:2016yue}, the value of the \gls{cp}-violating phase \cite{brahma2023probingdualnsicp}, the formation of black holes after a supernova explosion \cite{M_ller_2018}, and the hypothetical proton decay \cite{domingo2024novelprotondecaysignature}. The \gls{fd}, one of the components of \gls{dune}, will also utilise \gls{lartpc} technology \cite{Baller_2017}, which has successfully been used in a number of recent experiments \cite{icaruscollaboration2023icarusfermilabshortbaselineneutrino, MicroBooNE:2016pwy, DUNE:2020cqd, SBND:2020scp, Soderberg:2009qt}.  
Of particular interest in long-baseline neutrino oscillation experiments is the ratio of electron to muon neutrinos arriving at the detector, making the primary task in such experiments reconstructing and identifying events as muon neutrino or electron neutrino events. To do so, the spatial details of the event captured need to be understood thoroughly. An accurate classification of each of the signals received is thus paramount. This work will make use of the \gls{uboone} open dataset \cite{cerati2023microboonepublicdatasets}, but the longer-term objective of the project is to provide tools for processing data recorded by the \gls{dune}-\gls{fd}.

\leavevmode

Pandora \cite{Marshall:2015rfa,MicroBooNE:2017xvs,DUNE:2022wlc}, typically run as part of LArSoft \cite{Snider:2017wjd} workflows, is one of the most popular reconstruction frameworks for \gls{lartpc} experiments. It has a modular approach where each algorithm performs a specific, and typically small, task. In this way, it is easier to guarantee that each step is done correctly. One of the most important steps is to separate track and shower-like topologies at the hit level, a crucial task to perform particle identification and event reconstruction. In order to do so, Pandora makes extensive use of \gls{ml} techniques throughout the reconstruction chains. \gls{ml} and \gls{dl} have also been utilised in the \gls{uboone} experiment \cite{MicroBooNE:2021pvo}, and there are several proposed applications of these techniques in \gls{dune} \cite{DUNE:2020gpm, Machado:2020yxl, Kopp:2024lch, Moretti:2023rgd}. Different approaches to separate track-like particles travelling through dense electromagnetic showers have been explored \cite{Vergani:2022rtb,vergani_2024, DUNE:2022fiy, MicroBooNE:2020yze}, but it remains one of the many open reconstruction problems where an increase in the performance is still sought after. The presented work contributes to this important subroutine.

The exploration of ever more capable machine learning models and recent rapid advancements in quantum technologies led to the birth of \gls{qml}. One common critique of the current \gls{qml} research is that models which appear successful are often benchmarked on simple problems \cite{Bermejo:2024sgi,Bowles:2024fvp}. 
This study is interested in a real open problem in experimental neutrino physics, which remains unsolved by current machine learning methods.
Previous \gls{qml} uses in \gls{lartpc} experiments have been proposed in the context of event classification \cite{chen2021hybrid, chen2022quantum} as well as event generation \cite{delgado2024towards}.
Together with many other early studies investigating the use of quantum computers in \gls{hep}, the community is an active contributor to the development of quantum technologies whilst being posed to be their great beneficiary \cite{DiMeglio:2789149}. 

A high-level introduction to the field of \gls{qml} is given in Section~\ref{sec:qml}. In Section \ref{sec:datasets}, a detailed description of the \gls{lartpc} datasets used can be found. Section~\ref{sec:models} discusses model design concepts, and the quantum and classical architectures used are described in Section~\ref{sec:architectures}. In Section~\ref{sec:results}, one can find the results of applying the developed models to the problems.

\section{Quantum Machine Learning}\label{sec:qml}

Quantum machine learning has emerged as one of the most promising families of quantum algorithms. It has garnered significant attention due to the success and widespread adoption of classical machine learning and the compatibility of \gls{qml} methods with noisy intermediate-scale quantum (NISQ) \cite{Bharti_2022} devices. Combining these factors makes \gls{qml} a particularly attractive avenue for exploring quantum computational advantages in practical settings.

While the search for practical quantum advantages in \gls{qml} remains an active area of research, significant strides have been made theoretically \cite{Zheng_2023, Du_2020, Coyle_2020, J_ger_2023,sweke2025potentiallimitationsrandomfourier, Huang_2021}. These advances, combined with scrupulous dequantisation\footnote{Dequantisation refers to finding classical algorithms with complexity scaling matching a proposed quantum algorithm.} studies \cite{cerezo2023does}, continue to refine our understanding of the field's potential and its limitations. Empirically, small hints of potential advantage of quantum models appear in the form of better performance or similar performance with fewer parameters \cite{duffy2024unsupervisedbeyondstandardmodeleventdiscovery, PhysRevD.109.052002, tuysuz2024learning, chen2022quantum}. It is crucial, however, to state that benchmarking quantum models against classical ones is, in general, a difficult task \cite{Bowles:2024fvp}. We find ourselves in a time where classical simulations of \gls{qml} pipelines are prohibitively complex beyond $10$s of qubits but quantum hardware is not yet ready to train large quantum models, either. What we can do in the meantime is understand the models we work with well and test them on challenging, real-life problems.

A critical consideration in the design of quantum learning models is their inductive bias, which refers to the assumptions embedded in the model architecture that influence its generalisation capabilities. This principle is the focus of geometric \cite{ragone2023representationtheorygeometricquantum, Nguyen_2024,Meyer_2023} and scientific \cite{paine2023physicsinformedquantummachinelearning, Kyriienko_2024,williams2024addressingreadoutproblemquantum} \gls{qml}, where domain knowledge is incorporated into model design to enhance performance and interpretability.

The model used in this study is the well-known quanvolutional neural network which has been extended, for the first time, to include symmetries beyond translation. Quanvolutional neural networks allow one to study relatively large problems, compared to, for example, the \gls{qcnn} \cite{cong2019quantum}, as quantum computation is required only for a small subroutine of the full forward pass. The models process data mostly classically and use only small circuits on local patches of a data point. Details of the model can be found in Section \ref{sbsec:quanvolution}.

\section{Datasets}\label{sec:datasets}
Topologies seen in \gls{lartpc} images fall into two main categories: tracks and showers. Tracks are linear structures, whereas showers are dense collections of non-zero pixels. Showers are produced by electromagnetic cascades induced by electrons (e$^-$), positrons (e$^+$) and photons ($\gamma)$, whereas other particles such as muons ($\mu^\pm$), charged pions ($\pi^\pm$) and protons (p) produce tracks. Clear examples of tracks and showers can be seen in Figure~\ref{fig:event_examples}. 

This study uses two \gls{lartpc} datasets to test the efficacy of \gls{qml} in hit-based topology classification. The first is the openly available \gls{uboone} dataset \cite{Acciarri_2017, abratenko_2023_8370883}, which is described in detail in Section~\ref{sec:microboone}. The second is an original dataset produced specifically for this study. It is made up of events where a muon and an electron are produced at the same vertex but at various opening angles. It allows us to easily control the "difficulty" of classification of the events produced. Intuitively, the smaller the angle between the two particles, the ``denser" the events created are. Details of this dataset can be found in Section~\ref{sec:particle_bomb}.

\subsection{MicroBooNE open dataset}\label{sec:microboone}
\leavevmode
\begin{figure*}[t!]
    \centering
    \includegraphics[trim= 0cm 2cm 2cm 2cm,width=\linewidth]{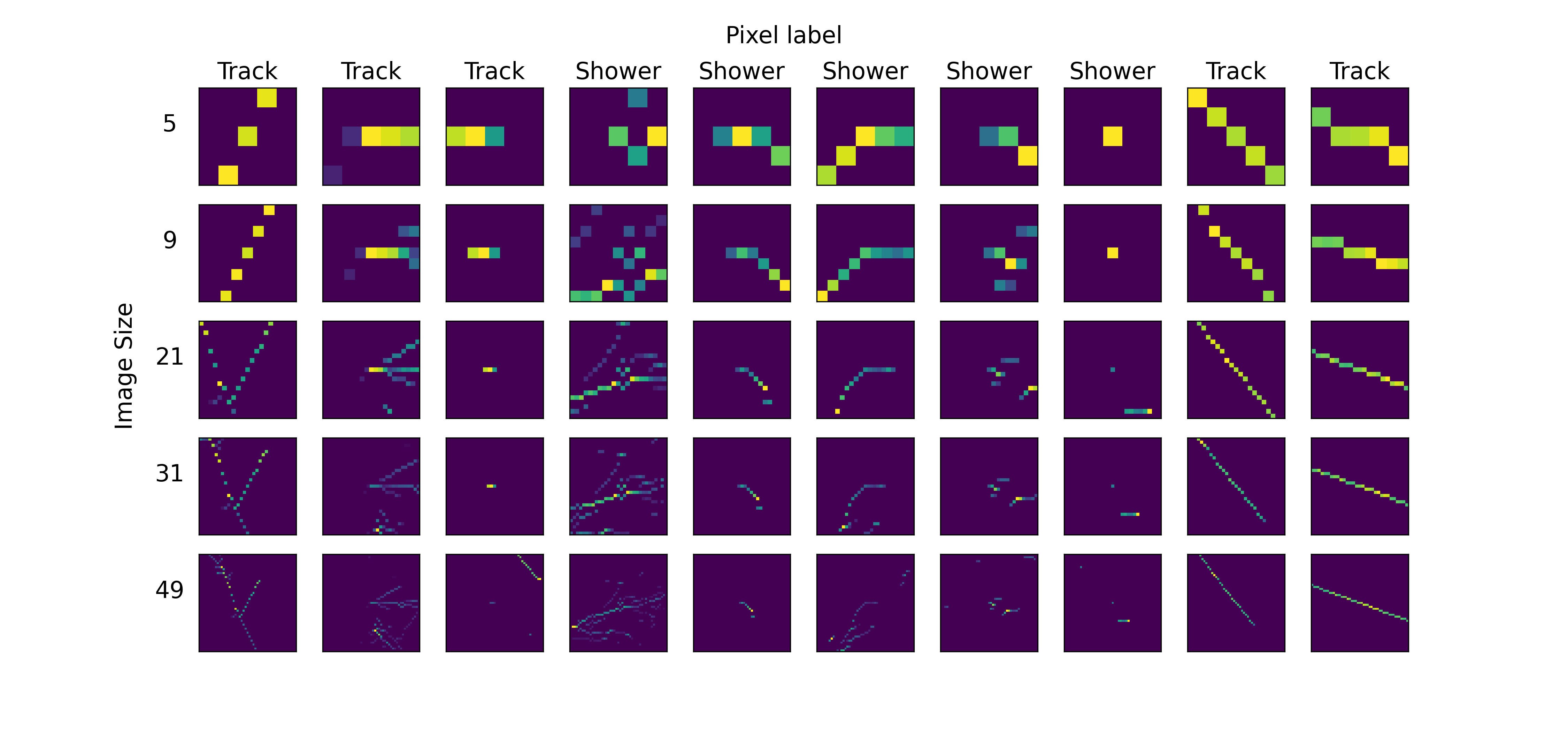}
    \caption{Randomly selected example event patches at various image sizes, categorised by pixel-level track and shower labels. Each row corresponds to a fixed patch size (given in pixels on the left), while each column shows the same event at different patch size scales. The label is determined by the classification of the central pixel as either track-like or shower-like. }
    \label{fig:microboone_example_datapoints}
\end{figure*}

This dataset, utilising the Booster Neutrino Beam (BNB) flux \cite{PhysRevD.79.072002}, consists of approximately 750,000 events simulated in the \gls{uboone} detector, featuring neutrino interactions producing electrons, photons, muons, charged pions, and protons and overlaid real cosmic rays. The \gls{uboone} experiment is a \gls{lartpc} neutrino detector, which operated as part of Fermilab's \gls{sbn} programme. The detector is positioned on the surface, which results in a significant background of cosmic-ray-induced particles, predominantly muons. Due to its long integration time, \gls{uboone} detector collects ionisation over an extended drift length, further amplifying cosmic-ray contamination. However, these cosmic-ray-induced particles have been removed for this study to ensure that the analysis focuses on distinguishing neutrino-induced track and shower events. Moreover, this is to align the study more closely with \gls{dune}, where the cosmic ray background will be negligible. No other cuts or modifications were applied to the dataset.

Each event in the dataset is represented as a set of three two-dimensional images, one for each wire readout plane: two induction planes (U and V), each consisting of $2400$ wires, and a collection plane (Y) with $3456$ wires. The induction planes are oriented at $\pm 60^{\circ}$ relative to the collection plane, whose wires run vertically. The wire spacing is $3$\,mm for all planes. In each image, the x-axis corresponds to the wire number, while the y-axis represents the drift time. Pixel intensity is proportional to the number of ionized electrons reaching that position and time, effectively encoding energy deposition information. The effective pixel resolution is $3.3$\,mm in the drift (y) direction and $3$\,mm in the wire (x) direction, as the waveform is integrated over six TPC time-ticks (3\,$\mathrm{\mu s}$).

The goal of this work is to separate track and shower topologies. To facilitate training, we subdivide event images into $N \times N$ pixel patches (See Figure~\ref{fig:microboone_example_datapoints}) and train on randomly shuffled patches across all the events. Mimicking the approach used in another study~\cite{Vergani:2022rtb}, a patch around each hit is built, with the chosen hit in the centre of the patch. Features of that patch are analysed to determine whether the central pixel is track or shower-like. By repeating this process for each hit in the image, the algorithm is able to characterise each hit in the image as track or shower-like. Based on the simulation labels, each patch is assigned a label based on the particle that contributes most to the intensity of the central pixel. While multiple charged particles can deposit charge at the same location, we adopt a single-particle labelling scheme, which has been shown to be beneficial for downstream tasks~\cite{Abratenko_2021}. At this initial stage of reconstruction, rather than identifying the specific particle responsible for the signal, each deposit is classified as track-like (label 0) or shower-like (label 1). This labelling scheme is well-motivated, as each charged particle produces either a track-like or shower-like signature.


\subsection{Neutrino-like dataset}\label{sec:particle_bomb}
\leavevmode
\begin{figure*}[t!]
    \centering   
    \includegraphics[width=0.32\linewidth]{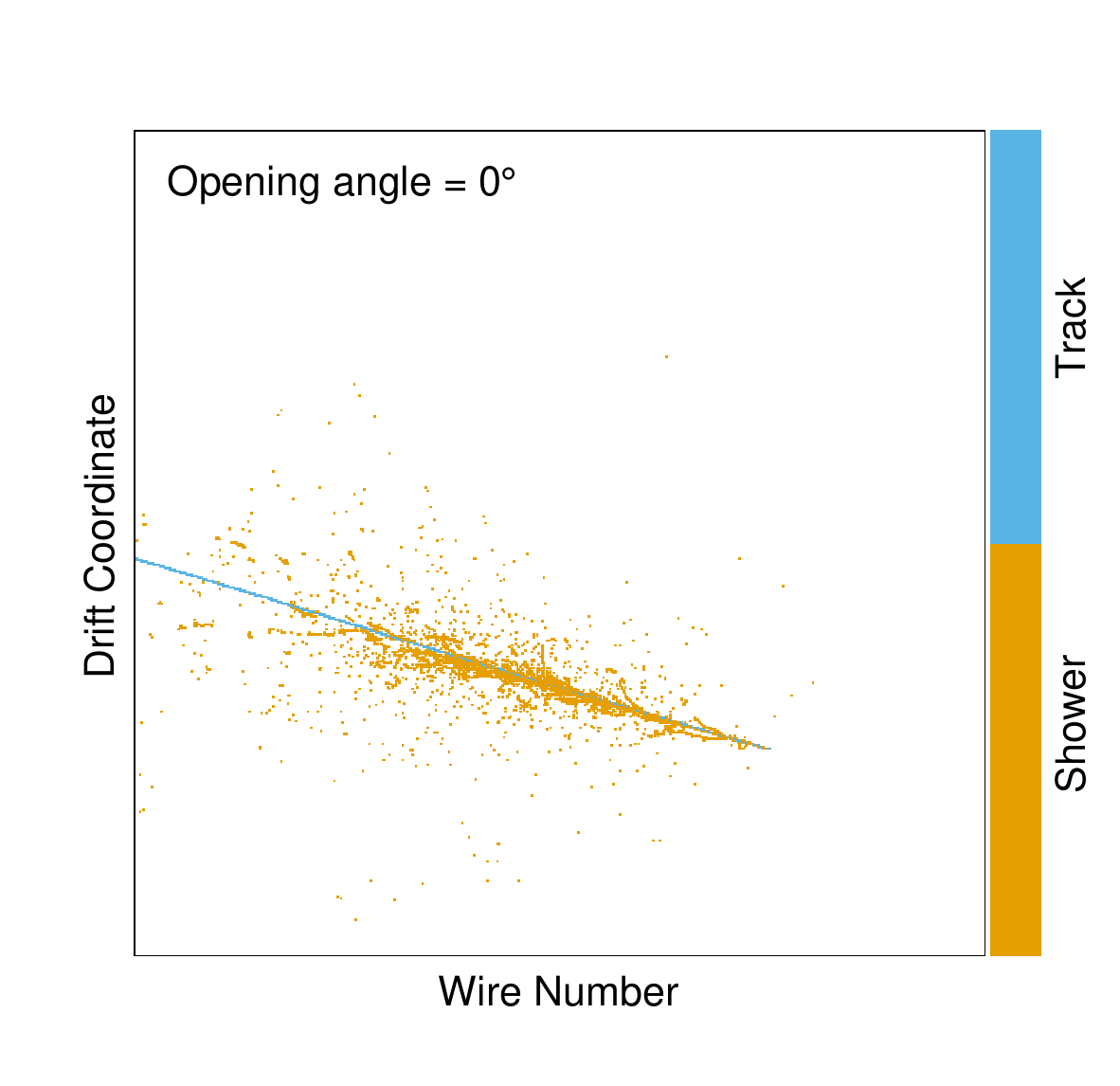}
    \includegraphics[width=0.32\linewidth]{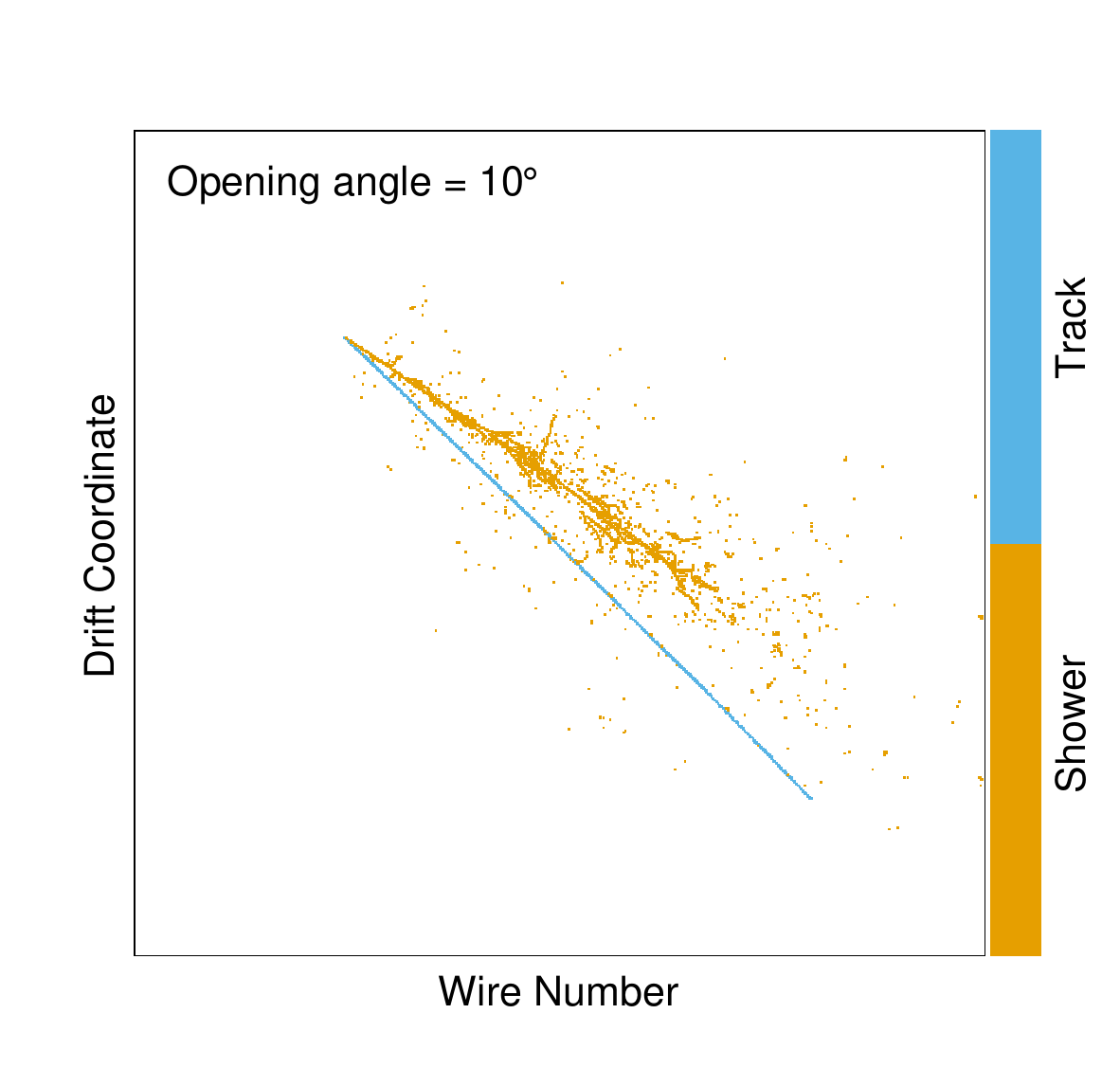}
    \includegraphics[width=0.32\linewidth]{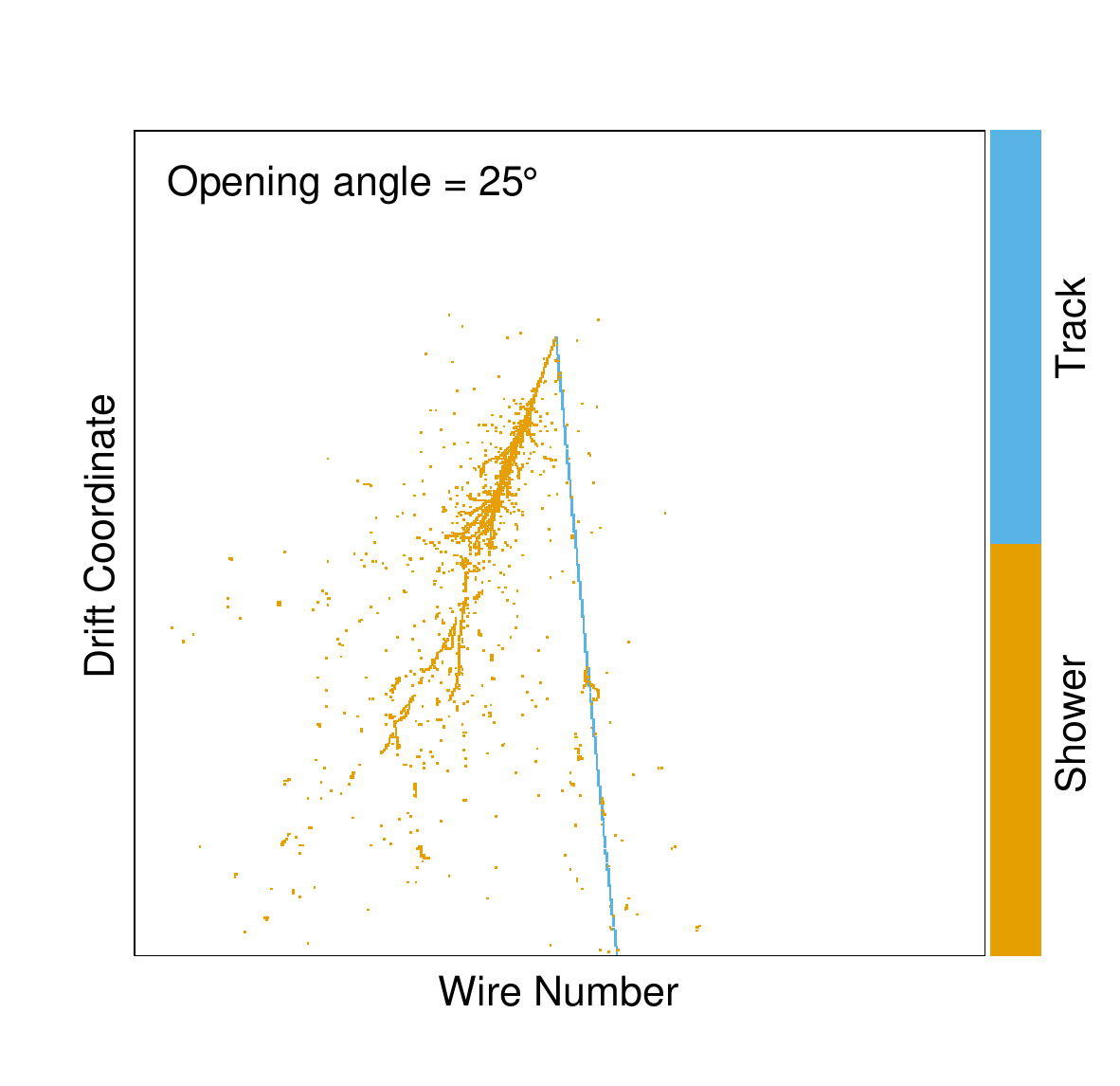}
    \caption{Three example events containing one electron and one muon produced at a common vertex position. The events are seen projected into one of the three readout views, and the projected opening angle is given. Shower-like energy deposits in the electron-initiated electromagnetic shower and the small ionisation electrons along the muon track are shown in orange, and the energy deposits coming directly from the muon are shown in blue.} 
\label{fig:event_examples}
\end{figure*}
The aim of this dataset is to create events with one shower-like and one track-like particle emanating from a common vertex with a variable opening angle between them. This was performed using LArSIMple, a simple \gls{lartpc} detector simulation~\cite{Chappell:2022yxd} based on Geant4 \texttt{v4\_11\_1\_p01ba}~\cite{Agostinelli:2002hh}. An electron was created with its direction randomly chosen from an isotropic distribution, and a muon was created within a cone of a variable opening angle around the direction of the electron. The produced muons and electrons have energies drawn from a flat distribution between 1.0\,GeV and 5.0\,GeV. Together, these two particles produce interactions topologies with one shower-like and one track-like object.

The particles are tracked through a cuboid detector filled with liquid argon and the dimensions in the $(x,y,z)$ directions are $5\,$m$\,\times\, 5\,$m$\,\times\, 5\,$m, where $z$ defines the beam direction, $y$ is vertical and $x$ is the drift direction. The simulation produces three-dimensional energy deposits within the detector volume that are converted into three images with height coming from the drift coordinate $x$ and width from one-dimensional projections of the $yz$ plane, similar to the three wire readout planes in the planned DUNE detectors~\cite{DUNE:2020txw}. These three views are referred to as $u$, $v$ and $w$ and are aligned at 35.9$^\circ$, -35.9$^\circ$ and 0$^\circ$ to the vertical, respectively.

Three example events are shown in Figure~\ref{fig:event_examples} with opening angles of $0^\circ$, $10^\circ$ and $25^\circ$ between the electron and muon, as seen in one of the three readout planes in the coordinates $\left(\textrm{wire},\,x\right)$. The shower-like activity in the events is shown in orange, and the track-like deposits are shown in blue. It is clear that as the opening angle increases, the amount of overlap between the track- and shower-like energy deposits decreases.

\section{Model design concepts}\label{sec:models}

The models used in this study are all based on \gls{cnn} architectures, which have achieved remarkable success in computer vision and beyond. \gls{cnn}s form the backbone of historic architectures such as AlexNet (2012) \cite{NIPS2012_c399862d}, ResNet (2015) \cite{he2015deepresiduallearningimage} and UNet (2015) \cite{10.1007/978-3-319-24574-4_28}. They also play a key role in modern generative models, such as Stable Diffusion (2022) \cite{Rombach2021HighResolutionIS}.
The convolutional layer uses a sliding filter which processes local windows of an image, sharing the same weights for all windows. This way it can effectively extract crucial features from the image.
Given that \gls{lartpc} events can be interpreted as a collection of images, as described in Section~\ref{sec:datasets}, \gls{cnn}s are naturally suited for their analysis. Numerous existing studies have successfully applied \gls{cnn}s to \gls{lartpc} data \cite{Domin__2020,Liu:2020pzv,DUNE:2020gpm,Moretti:2023rgd,MicroBooNE:2021pvo,Acciarri_2017,DUNE:2022fiy}.

One of the main properties of a convolutional layer is translation equivariance; the property of ``respecting" translation symmetries of the input data. Section~\ref{sbsec:eq_convolution} explores this property in more detail and discusses how convolutions can be extended to respect symmetries beyond translation.

\begin{figure*}[t!]
    \centering
    \includegraphics[trim= 0cm 4cm 0 4cm, width=\linewidth]{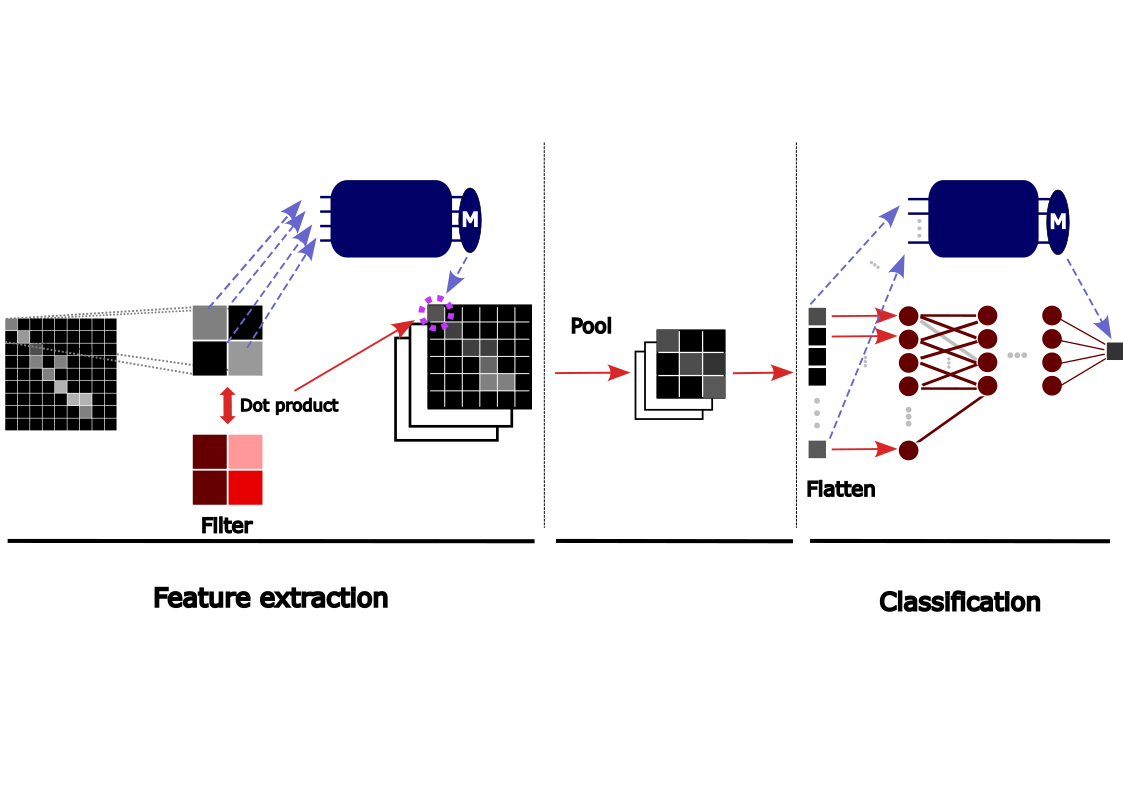}
    \caption{A convolutional neural network and its extension to a quanvolution network. Data (greyscale) can be processed classically (red) or using quantum circuits (blue) where indicated. Quantum circuits can be used as a replacement for dot products with classical filters or classical classifier networks (See Section \ref{sec:models} and Appendix \ref{sec:Appendix_mis_quanv}). $M$ indicates a measurement whose expectation value is the output of the circuit.}
    \label{fig:quanvolution}
\end{figure*}

In this study, quantum circuits are introduced as replacements for specific components of a standard \gls{cnn} architecture, resulting in a quanvolutional network. The quanvolutional layer, detailed in Section~\ref{sbsec:quanvolution}, modifies classical convolution while inheriting the symmetries of any convolutional layer. After a series of convolutional or quanvolutional operations, the generated feature maps are passed to a classifier, which may be either classical (a \gls{mlp}) or quantum (a \gls{pqc}). Section~\ref{sbsc:invariant_quantum_classifier} discusses the use of an equivariant quantum classifier for this task. Pooling layers may be introduced when dimensionality reduction is needed (See Figure \ref{fig:quanvolution}). 

\leavevmode

\subsection{Convolution and its equivariance}
\label{sbsec:eq_convolution}
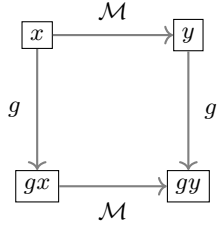
\begin{figure}
\centering
    \begin{tikzpicture}
    \node[style=rectangle, draw=black, fill=white](x) at (-1,1) {$x$};
    \node[style=rectangle, draw=black, fill=white] (y) at (1,1) {$y$};
    \node[style=rectangle, draw=black, fill=white] (gx) at (-1,-1) {$gx$};
    \node[style=rectangle, draw=black, fill=white] (gy) at (1,-1) {$gy$};
    
    \draw[gray, thick, ->] (x) -- (gx);
    \draw[gray, thick, ->] (y) -- (gy);
    \draw[gray, thick, ->] (x) -- (y);
    \draw[gray, thick, ->] (gx) -- (gy);

    \path (x) -- (y) coordinate[midway] (midpoint_up);
    \path (gx) -- (gy) coordinate[midway] (midpoint_down);
    \path (x) -- (gx) coordinate[midway] (midpoint_left);
    \path (y) -- (gy) coordinate[midway] (midpoint_right);
    \node[above=0.1cm of midpoint_up] (M) {$\mathcal{M}$};
    \node[below=0.1cm of midpoint_down] (M) {$\mathcal{M}$};
    \node[left=0.1cm of midpoint_left] (M) {$g$};
    \node[right=0.1cm of midpoint_right] (M) {$g$};
    
    \end{tikzpicture}
    \caption{Equivariant map $\mathcal{M}$ and an element $g$ of a group $G$ acting on elements $x$, $y$ of two spaces $X$, $Y$, respectively.}
    \label{fig:equivariance_general}
\end{figure}

Equivariance is a property of a map $\mathcal{M}: X\rightarrow Y$ between two spaces (formally, G-sets) satisfied when \cite{adhikari2013basic}:
\begin{equation}
    \mathcal{M}(gx)= g\mathcal{M}(x) \quad \forall g \in G,\; \forall x \in X.
\end{equation}
This can be interpreted as: acting with a group element before acting with the map results in the same object in $Y$ as first mapping to $Y$ and then acting with the group element (See Figure \ref{fig:equivariance_general}).

\leavevmode

The convolution layers of a \gls{cnn} are translation-equivariant maps. This means that if the original input becomes shifted, the output of the layer will be shifted accordingly. Mathematically, a convolution\footnote{Technically, this is a closely related function called cross-correlation, but the term convolution is prevalent in ML literature.} can be expressed as~\cite{worrall2018cubenet, cohen2019general}:

\begin{equation}
    [F\star W](g) = \sum_{h\in {G}} W(g^{-1}h)F(h),
    \label{eq:convolution}
\end{equation}
where $g \in G_{out}$ are elements of the group the output is defined over, $h \in G$ are the elements of the group the input is defined over (can be different to $G_{out}$), $W$ is the filter and $F$ is the feature map (original data point in the case of the first convolution layer). Having defined the convolution this way, we can express the most common 2-D convolution by taking $G = G_{out}=(\mathbb{Z}^2, +)$, the (pixelised) translation group.

For tasks dealing with images, of interest are convolutions defined over roto-translation groups; subgroups of the Euclidean group $E(2)$ \cite{weiler2019general}. The Euclidean group can be expressed as a semidirect product of the translation group $(\mathbb{R}^2,+)$ and the orthogonal group $O(2)$ which includes all continuous translations and reflections. For many image datasets, their crucial features do not depend on the orientation of the object in a picture. Here, we hypothesise that orientation does not play a role in the topology of a track or shower.

For this study, we shrink the $O(2)$ group to just the $C_4$ group (the four $90^{\circ}$ rotations) for simplicity. This results in an overall  $(\mathbb{Z}^2, +) \rtimes C_4$ group of interest. This can be implemented by having each pixel contain a vector of four values, one for each element of $C_4$, and correctly applying group elements to the filters in Equation \ref{eq:convolution}.

We note that this is the first investigation into the use of Euclidean symmetries on \gls{lartpc} images.

\subsection{Quanvolution and its equivariance}
\label{sbsec:quanvolution}
The quanvolutional neural network has been proposed in \cite{Henderson2019QuanvolutionalNN} and \cite{Liu_2021} as a quantum-enhanced extension to the convolutional neural network. 

We first note a common misinterpretation of the quanvolution: this architecture is often presented as using quantum circuits as ``quantum filters"\footnote{A convolution filter is often also referred to as a \textit{kernel}.} performing a convolution \cite{Liu_2021,Henderson2019QuanvolutionalNN, chen2022quantum, pennylane_quanv}. 
Appendix \ref{sec:Appendix_mis_quanv} shows that this interpretation is misleading and highlights a more subtle but crucial relationship between the quanvolution and convolution. This new interpretation can be used to show how quanvolutions obtain their equivariance property.

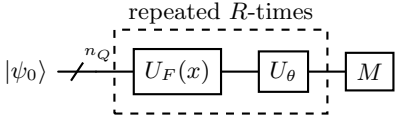
\begin{figure}
    \centering

    \begin{quantikz}
         \lstick{\ket{\psi_0}}  &  \qwbundle{n_Q} & \gate[1]{U_F(x)} \gategroup[1, steps = 2, style={dashed}]{repeated $R$-times} & \gate[1]{U_{\theta}} & \gate[1]{M} 
    \end{quantikz}
    \caption{General $n_Q$-qubit reuploader circuit architecture used for all circuits in this work. $U_F(x)$ is a data-encoding unitary for a datapoint $x$. $U_{\theta}$ is a trainable ansatz which does not depend on data. $M$ denotes a chosen measurement unitary whose expectation value is used as the output of the trainable circuit. Exact choices for $U_F(x)$,  $U_\theta$,  $M$ are discussed in the relevant sections of the text.}
    \label{fig:reuploader_circuit}
\end{figure}

The quanvolution layer takes, as input, a feature map $F$. Windows of the feature map (like those in a convolution) are embedded in a unitary $U_{F,W}$ on the Hilbert space of a \gls{pqc}. The number of qubits $n_Q$ used is ordinarily equal to the number of pixels in the window. The full circuit processes the pixels contained in the window and returns a real number as output. This is usually obtained via an expectation value of a chosen observable $M$. This real number is used to define the feature map of the following layer. The feature extraction part of Figure \ref{fig:quanvolution} represents this visually.
In this work, the quanvolution circuits used are reuploader circuits of the form:
\begin{equation}
\label{eq:reuploader}
    F_{out}(g) = \langle \psi_0|(U_{F,W}^{\dagger}U_\theta^{\dagger}(g))^{R}|M|(U_{\theta}U_{F,W}(g))^{R}|\psi_0 \rangle,
\end{equation}
where $|\psi_0\rangle$ is some initial state of the qubits, taken in this study to be the $|0\rangle^{\otimes n_Q}$, $U_\theta$ is an ansatz - a unitary with trainable parameters and $R$ is the number of times the $U_\theta U_{F,W}(g)$-block is repeated. A general circuit of this kind is pictured in Figure \ref{fig:reuploader_circuit}.

$U_{F,W}$ used was the rotation embedding 
\begin{equation}
\label{eq:rot_embedding}
    U_{F,W}(g) = \prod_{i\in[n_Q]} RX_{i}(F_{W(g),i}),
\end{equation} 
where $RP$ denotes a Pauli $P$ rotation, $P\in{X,Y,Z}$ and $F_{W(g),i}$ being the value of the feature map (pixel) assigned to qubit $i$ in the window of $g$. $U_\theta$ was a nearest-neighbour entangling ansatz with two trainable parameters per qubit, per entanglement block layer, 
\begin{equation}
\label{eq:ansatz}
\begin{split}
  U_\theta &= \prod_{l \in [L]}RZ_{n_Q}(\theta_{2,l,n_Q})RY_{n_Q}(\theta_{1,l,n_Q}) \\
  &\prod_{i \in [n_Q-1]}CX_{i,i+1}RZ_{i}(\theta_{2,l,i})RY_{i}(\theta_{1,l,i}),
  \end{split}
\end{equation}
where $CX_{i,j}$ is a controlled X (also called a controlled NOT) gate with $i$ being the control and $j$ the controlled qubit and $L$ is a hyperparameter denoting the number of layers of the ansatz (See Table~\ref{tab:hyperparams}).

A $G$-equivariant quanvolution can be achieved in analogy to the standard quanvolution. That is, every dot product between a feature map and a filter of the equivariant convolution described in Section~\ref{sbsec:eq_convolution} is replaced with a pass of the feature map through a quantum circuit, as described earlier in this Section. We note that the equivariance of this layer comes purely from its classical construction and the use of identical circuit structure for each group element $g$. Appendix~\ref{sec:Appendix_mis_quanv} shows how a quanvolution inherits its equivariance from a convolution. In other words, the individual circuits used do not need to be equivariant maps. This is unlike what will be discussed in Section~\ref{sbsc:invariant_quantum_classifier}, where the invariant quantum classifier is constructed using methods from geometric quantum machine learning.

As mentioned earlier, it is customary to use quanvolution circuits with the same number of qubits as the number of pixels in the sliding window of the quanvolution. There is, however, no need to adhere to this restriction. As long as the quanvolution circuits used depend on a given group element solely via their embedding, any size circuit can be used. Growing the circuit to many tens of qubits will likely be necessary for it to be intractable classically.

Additionally, whilst small filter sizes are commonly used in \gls{cnn}s, some notable results from the literature show that large filters can achieve competitive performance \cite{ding2022scaling, liu2022more}. Future studies should investigate increasing the size of the sliding window in quanvolutions. Depending on the ansatz used, this could quickly become intractable to known classical simulation techniques.

\leavevmode

\subsection{Invariant quantum classifier}
\label{sbsc:invariant_quantum_classifier}
Equation \ref{eq:reuploader} is, at its core, a description of a quantum regression model. If we forget about the group structure required for quanvolutions, we can re-write it more generally, for any input data point $x$ as:
\begin{equation}
    f(x) = \langle \psi_0|(U_{x}^{\dagger}U_\theta^{\dagger})^{R}|M|(U_{\theta}U_x^{\dagger})^{R}|\psi_0 \rangle.
\end{equation}
The output such a model is a real number, which can be turned into a binary label by choosing a threshold, e.g. $0$. This is how the classification part of Figure \ref{fig:quanvolution} can be realised by a quantum circuit.

Geometric quantum learning has bloomed in the recent years. Theoretical results show that some geometric models have desirable properties \cite{Schatzki_2024} and have been proposed as a strong candidate for potential quantum speedup  \cite{Zheng_2023}. Given a group, finding an equivariant ansatz can be performed in a variety of ways \cite{Nguyen_2024}. A quantum classifier acting on classical data can be made \textit{invariant} if the embedding, the ansatz and the measurement are equivariant \cite{ragone2023representationtheorygeometricquantum, Nguyen_2024}.

The classifier circuits in this work take as input the last layer of the feature extraction part of the network (See Figure~\ref{fig:quanvolution}). Preserving rotation equivariance between the two parts of the network can be ensured by averaging along the $C_4$ dimension as well as along the regular convolution channels, resulting in a single image representing all the information contained in the last feature extraction layer. Other schemes for this procedure might be possible and could be explored in the future.

The invariant circuits used as classifiers in the study also follow the general structure of Figure \ref{fig:reuploader_circuit}. $U_x$ is chosen again to be the rotation embedding described in Equation \ref{eq:rot_embedding}, this time with the whole feature map  image being passed (no sliding window necessary). This embedding is equivariant to $C_4$ rotations. The ansatz $U_\theta$ is constructed using equivariant gates according to a scheme detailed in Appendix \ref{sec:Appendix_geo_class} and $M$ is a $Z$-basis measurement of the central qubit, also equivariant to $C_4$.

Given the ansatz and hyperparameters used, the number of parameters in this layer is much smaller (on the order of $10$) than those used by the \gls{mlp}s of the other models (on the order of $10^3$).

\section{Model architectures}\label{sec:architectures}

This section details how the concepts from the previous Section have been combined to design the specific architectures used on the \gls{lartpc} datasets.

We consider the \gls{cnn} to be made up of two modules: the feature extraction module (with optional pooling after each layer) and the classification module. To help keep track of the different variants of the models, we refer to them by the symmetry (equivariance) status $S$ and type $T$ of their modules, where $S \in \{E, NE\}$ and $T \in \{Q, C\}$.  $E$ denotes `equivariant', $NE$ - `non-equivariant', $Q$ - `quantum', $C$ - `classical'. Together, a given architecture can be referred to as $TS \rightarrow TS$. Table \ref{tab:models} details all the models used.
\renewcommand{\arraystretch}{1.1}
\begin{table}[h!]
    \centering
    \begin{tabular}{c|c|c|c}
         \textbf{Model} & \makecell{Feature \\ extraction} & Classification & \# params \\ 
         \hline 
         \textbf{Deep CNN}&$CNE$& $CNE$ & $10^5$ \\ 
         \textbf{CNN}&$CNE$& $CNE$ & $10^3$ \\ 
         \textbf{Quanv}&$QNE$& $CNE$ & $10^3$ \\ \hline
         \textbf{Deep GCNN}&$CE$& $CNE$ & $10^5$ \\ 
         \textbf{GQuanv}&$QE$& $CNE$ & $10^3$ \\ 
         \textbf{InvQuanv}&$QE$& $QE$ & $10^2$ \\ 
    \end{tabular}
    \caption{Summary of the architectures used in the study using the nomenclature introduced in the text. Number of parameters is given as an order of magnitude estimate as the choice of hyperparameters impacts the exact number.
    Top three models have no symmetries beyond translations. The remaining are equivariant to $C_4$ rotations at least in the feature extraction part.}
    \label{tab:models}
\end{table}

\subsection{Classical benchmarks}

Three classical benchmark models are used in this study:  

\begin{itemize}
    \item \textbf{Deep CNN ($CNE \rightarrow CNE$)} – A six-layer convolutional neural network with $3 \times 3$ filters and output channels $[16,\,16,\,32,\,32,\,64,\,64]$, in each layer respectively.  
    \item \textbf{CNN ($CNE \rightarrow CNE$)} – A compact convolutional network with $2$ layers of $2 \times 2$ filters and three output channels per layer.
    
    \item \textbf{Deep GCNN ($CE \rightarrow CNE$)} – A deep \gls{cnn} that incorporates $C_4$ group convolutions, introducing rotational symmetry while maintaining the same overall architecture as the deep \gls{cnn}.  
\end{itemize}  

All models use an \gls{mlp} with two hidden layers as the final classifier. The shape of the hidden layers was a hyperparameter (See Appendix \ref{sec:Appendix_hypopt})

\subsection{Quantum-enhanced architectures}
Three quanvolutional networks with $2\times2$ quanvolution windows are implemented:

\begin{itemize}
    \item \textbf{Quanv $QNE \rightarrow CNE$} – Two quanvolutional layers, each with three output channels, followed by an \gls{mlp}.
    \item \textbf{GQuanv $QE \rightarrow CNE$} – Two $C_4$-symmetric quanvolutional layers, each with three output channels. Classification with an \gls{mlp}.
    \item \textbf{InvQuanv $QE \rightarrow QE$} – Two $C_4$-symmetric quanvolutional layers, each with three output channels. Classification with a $C_4$-symmetric quantum classifier using an equivariant embedding, equivariant reuploader circuits and an equivariant measurement.
\end{itemize}

Quanv is a standard quanvolutional network, utilising only the $(Z_2, +)$ group. The other two are rotation-aware. Feature extraction in both is achieved with a $C_4$-symmetric quanvolution described in Section~\ref{sbsec:quanvolution}. They differ only in their classifier module. The first uses a classical \gls{mlp} and the other, a quantum, symmetry-aware classifier discussed in Section~\ref{sbsc:invariant_quantum_classifier}.. The second model is a fully \textit{invariant} classifier. This means that two images which differ by a $C_4$ rotation will result in the same output, leading necessarily to the same label.

\section{Results}\label{sec:results}
For clarity, all plots in this section contain only the best performing quantum, classical and deep classical models. Figures with the performance of all models discussed can be seen in Appendix \ref{sec:Appendix_full_plots}.
\subsection{Performance on the MicroBooNE open dataset}

In this section, we evaluate the performance of our proposed models on the open \gls{uboone} dataset. Specifically, we investigate how the spatial context surrounding a classified pixel influences the model's performance. Figure \ref{fig:microboone_example_datapoints} illustrates examples of the same pixel within an event as part of progressively larger patches.

In our framework, a patch refers to the local neighbourhood of a pixel that is provided as input to the model. While the model ultimately classifies only the central pixel (as either ``track" or ``shower"), it utilizes the surrounding information within the patch to make this determination. For instance, if a pixel lies within a straight, connected path and exhibits consistent energy signatures, the model is likely to classify it as part of a track.

Incorporating a larger field of view offers contextual information about the broader topology, aiding classification. However, excessively large patches may introduce unrelated or irrelevant structures, potentially degrading model performance by introducing noise. The optimal patch size for LArTPC pixel classification remains an open question. In this work, we contribute to this ongoing investigation by evaluating model performance on the simplified \gls{uboone} dataset described in Section~\ref{sec:microboone}.

\begin{figure}[h!]
    \centering
    \includegraphics[width = \linewidth]{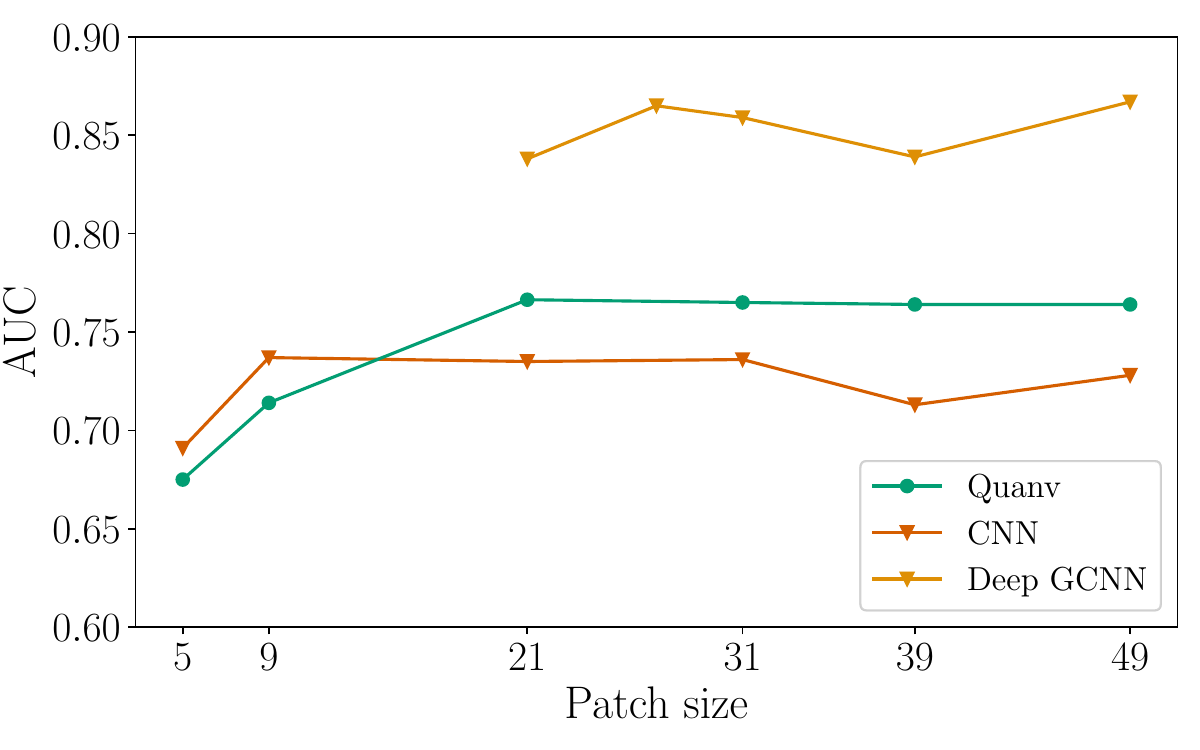}
    \caption{Receiver operating characteristic area under curve (ROC AUC) for best performing architectures used in this study with increasing pixel patch size. The deep model was not defined for small patches as no padding was used and it would reduce the size of feature maps to $1 \times 1$ before classification. Models were trained on $500$ patches.}
    \label{fig:patch_size_plot}
\end{figure}

Figure \ref{fig:patch_size_plot} presents the performance of the best performing quantum, classical and deep classical architectures described in Section~\ref{sec:models} across different patch sizes. Notably, larger models exhibit the ability to process increasingly large patches effectively—the more contextual information they receive, the better their classification performance. Conversely, smaller models demonstrate a decline in performance when presented with patches that are too large, likely due to an inability to process the additional information effectively.

Quantum models demonstrate superior performance compared to their classical counterparts across multiple patch sizes when constrained to a similar number of trainable parameters and with an identical number of filters. A $100$-fold increase in the number of parameters allows classical models to surpass the performance of the quantum-enhanced architectures.

It is unclear whether the inclusion of $C_4$ geometry in the models can be of undeniable benefit. Although among the deep models, the symmetric model showcased a consistently slightly better performance, among the quantum models, this bias shows no real advantage. A study incorporating larger rotation groups into the models is paramount.

\leavevmode

A key limitation of large-scale studies involving the simulation of quantum devices on classical hardware is the substantial computational cost in terms of time and RAM. In our study, performing hyperparameter optimization for selecting the best model (details in Appendix \ref{sec:Appendix_hypopt}) at each patch size (Figure \ref{fig:patch_size_plot}) requires multiple days per patch size. This computational overhead imposes constraints on the volume of training data and number of filters that can be incorporated into the optimization process.

To further investigate the learning capabilities of our models, we analyse their performance as a function of training dataset size. Specifically, we evaluate the best-performing hyperparameter configurations for models found in the previous analysis. Using a patch size of $21$, we check their performance across training set sizes ranging from $100$ to $1000$ data points. The metrics for each model are reported by averaging the $10$ best-performing instances from a run, using hundreds of random seeds for each training size. We report the mean, standard deviation, and maximum performance. The primary goal of this study is to assess the extent to which models can generalize in a data-constrained setting.

\begin{figure}[h!]
    \centering
    \includegraphics[width = \linewidth]{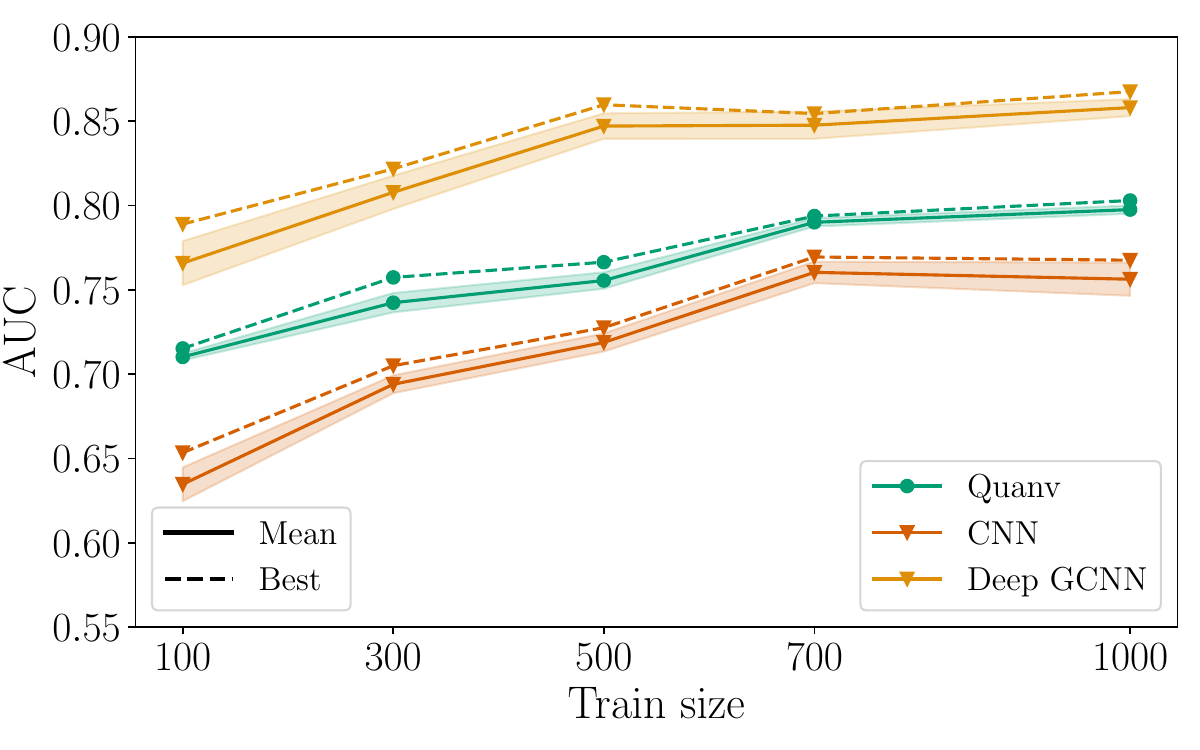}
    \caption{Learning capability of models used in this study. Vertical axis is the receiver operating characteristic area under curve (ROC AUC) and the horizontal axis is a training size (number of patches). Patch size was kept at $21$.}
    \label{fig:train_size_plot}
\end{figure}

The results of this experiment are presented in Figure \ref{fig:train_size_plot}. The overall trend confirms the hierarchy of the quantum, classical and deep classical architectures in the low-data regime we are able to probe.

The greater performance of the quanvolutional \gls{nn} compared to the shallow \gls{cnn} suggests that the representations learned by the quanvolutional \gls{nn} are richer and more informative. This enables better feature extraction for the final classification layers,  a trend which remains even as we increase the amount of training data. This suggests that (classical) quantum-inspired or actual quantum-enhanced architectures may offer competitive feature representations. However, further investigations into their scalability with larger datasets are warranted. 
\leavevmode

\subsection{Performance on the neutrino-like dataset}
\label{sec:results_particle_bomb}

As described in Section~\ref{sec:particle_bomb}, events in this dataset are characterized by a simple yet informative parameter — the angle between the two created particles. This angle serves as a proxy for the classification difficulty of individual pixels: smaller angles correspond to more challenging classification tasks due to increased spatial overlap and denser regions of energy deposition.

Understanding how models perform in high-density regions is crucial for future LArTPC experiments, as these regions present a major bottleneck for accurate event reconstruction. To systematically investigate how model performance scales with complexity, we divide the dataset into three categories based on the inter-particle angle:
\begin{itemize}
    \item \textbf{Easy}: $15^{\circ}-180^{\circ}$ (widely separated particles)

    \item \textbf{Medium}: $5^{\circ}-15^{\circ}$ (moderate overlap)

    \item \textbf{Hard}: $0^{\circ}-5^{\circ}$ (significant overlap and the most complex topology)
\end{itemize}

\begin{figure}[h!]
    \centering
    \includegraphics[width = \linewidth]{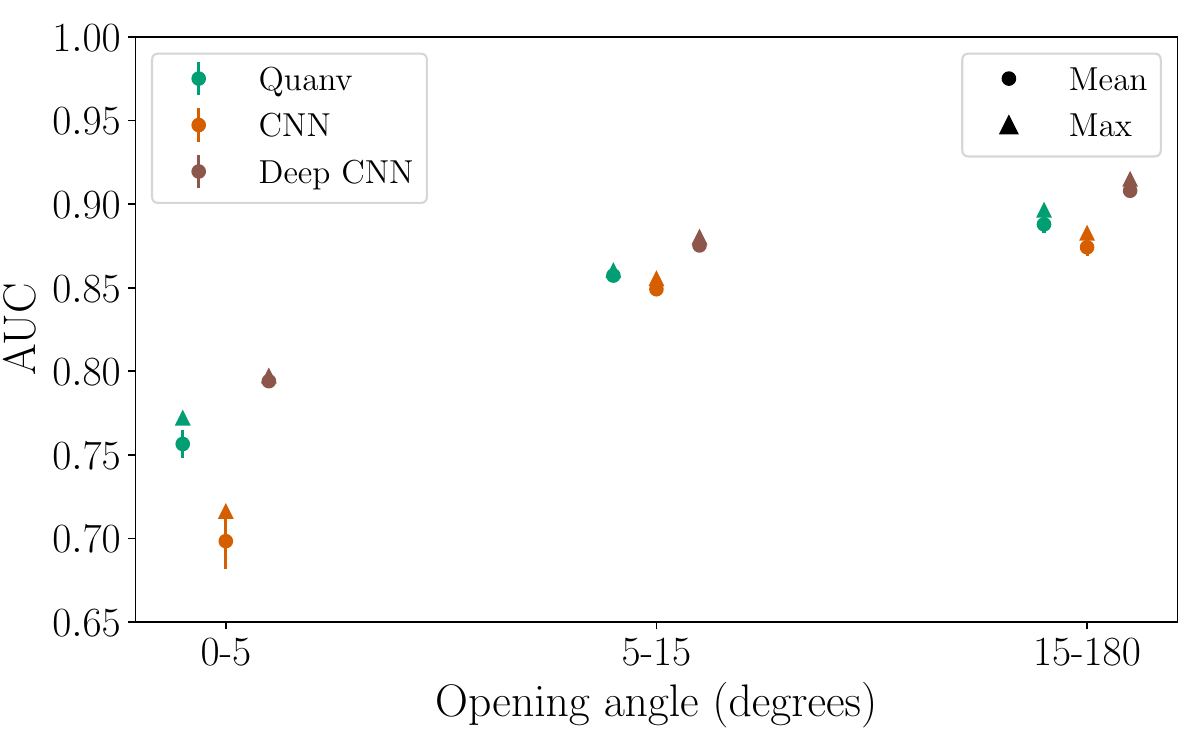}
    \caption{Performance in different environments of selected models in this study. Vertical axis is the receiver operating characteristic area under curve (ROC AUC) and the horizontal axis is the opening angle between the electron and muon particles. The patch size was kept at $21$.}
    \label{fig:particle_bomb_plot}
\end{figure}

We note that these bins were chosen empirically based on visual inspection of the events not on a detailed examination of the complexity of individual patches. One could devise more precise metrics by quantifying the amount of ``shower" pixels in a ``track" patch, for example. Results in Section \ref{sec:results_particle_bomb} confirm the average-case hardness of the three datasets created.
The performance of the models across these three categories is summarized in Figure \ref{fig:particle_bomb_plot}. A clear pattern emerges when comparing the shallow \gls{cnn} to the quanvolutional \gls{nn}: while the quanvolutional \gls{nn} consistently outperforms the shallow \gls{cnn} across all levels of difficulty, the largest performance gap appears in the hardest category ($0^{\circ}-5^{\circ}$). This suggests that the quanvolutional \gls{nn} is better equipped to resolve complex topologies and extract meaningful features, a crucial property when dealing with busy detector regions. We can potentially attribute this to the richer feature maps produced by the quanvolutional \gls{nn}.\

However, we also observe that larger classical \gls{cnn}s—with significantly more filters and overall capacity—continue to outperform the quanvolutional \gls{nn} across all difficulty levels. Interestingly, it is the non-equivariant architecture which, in general, prevails. Whilst each of the events in this dataset has a sense of direction (most patches will have structures oriented along the direction of the muon), which could render the addition of symmetries unhelpful, the patches used for training and testing are sampled from $100$ different events, where that direction is isotropic.

\leavevmode

\section{Conclusion}\label{sec:conclusion}

This paper contributes to the fields of experimental neutrino physics and quantum machine learning. In the context of particle physics, we address crucial open questions in \gls{ml}-driven pixel-level \gls{lartpc} classification. We examine the effects of patch size on the performance of the models, the benefits of including rotational symmetries in their design and begin to study their behaviour in dense environments. In the quantum machine learning domain, we point out a common misconception about a popular model, define it in a rigorous way based on its connection to the convolution layer, and show, for the first time, how to design it such that it respects symmetries beyond translation.

Results show that quantum-enhanced models are competitive with fully classical ones. For a similar number of parameters, quantum models consistently achieve better results. They remain outperformed by classical models with $10$ times more parameters, however. This trend holds for all the analyses implemented within this work.

Using the MicroBooNE data, it has been shown that large models can utilise large patches for pixel classification, whilst the amount of information contained in these large patches becomes confusing for smaller models. These find optimal performance at a patch size of $21$. An obvious question follows: can the large models benefit from even larger patches?

One downside of our study is that we operate in a very low-data regime, showing results for models trained on up to only $1000$ samples. Given the amount of data available in \gls{lartpc} experiments and \gls{hep} experiments in general, the use of more efficient \gls{qml} pipeline techniques becomes paramount. Future studies should strive to reduce the computational training and test time by leveraging GPUs, just-in-time compilation or simulations beyond state-vector-methods.

Implementing a custom ``neutrino-like" dataset, we defined a single parameter (angle between the two produced particles) as a proxy to denote the difficulty of classifying patches from a given event. Recognising a similar way to categorise pixels in actual experiments could become a highly beneficial tool for analysing the performance of the implemented classifiers.

The inclusion of $C_4$ symmetry in the models does not guarantee a substantial advantage. No advantage is seen within the (small) quanvolution models - the non-symmetric quanvolution performs best in almost all scenarios.  A small advantage of the (large) symmetric classial model over its non-symmetric counterpart can be seen in the MicroBooNE dataset but not in the neutrino-like dataset. A study varying the symmetry group (adding reflections, probing $C_n$ with $n\neq4$, or going to the continuous $SO(2)$ regime) is needed to determine which (if any) provides the best inductive bias for models dealing with \gls{lartpc} images. The case for discrete groups in quanvolutions can be explored using methods described in this work.

An obvious direction for future study is an extension to larger subgroups of the Euclidean group. 
We note also that some potentially interesting combinations of the feature extraction and classification modules have not been explored. Namely,
$CE\rightarrow CE$, $QE\rightarrow CE$ and $CE\rightarrow QE$. These would be fully invariant classifiers. Contrasting the properties and performance of all the combinations of the feature extraction and classification modules could further our understanding of any potential advantages offered by quantum elements in the architecture, as well as the inclusion of symmetry in the model.

Another interesting continuation of this study is the separation of similar shower-like deposits coming from different particles. In dense regions, different interactions can produce shower-like topologies that are very close to each other or even overlap, as is the case with the decay photons from $\pi^0$ mesons. As it remains crucial for a correct reconstruction to separate each deposit at the pixel level, work in that direction would likely be beneficial.

A \gls{qml} approach, such as the one proposed for this work, could be in the future implemented inside pattern recognition tools such as Pandora. Given Pandora's modular and multi-algorithmic nature, \gls{qml} algorithms could be used alongside classical \gls{cnn}s. Depending on the performance, \gls{qml} could be deployed to solve a particular problem where classical computing cannot. This would be particularly valuable with data coming from big detectors such as the \gls{dune}-\gls{fd}, where we expect a very high level of complexity and a hybrid approach in reconstruction could be very valuable.






\section*{Acknowledgements}
MJ thanks Peter T J Bradshaw for valuable discussions.
We acknowledge the MicroBooNE Collaboration for making publicly available the data sets~\cite{abratenko_2023_8370883} employed in this work. These data sets consist of simulated neutrino interactions from the Booster Neutrino Beamline overlaid on top of cosmic data collected with the MicroBooNE detector~\cite{Acciarri_2017}. The datasets were produced using the MicroBooNE Genie Tune \cite{PhysRevD.105.072001}.










\appendix
\section{Results for all models}
\label{sec:Appendix_full_plots}
This section contains full versions of Figures~\ref{fig:patch_size_plot}-\ref{fig:particle_bomb_plot}.
\leavevmode
\begin{figure}[h!]
    \centering
    \includegraphics[width = \linewidth]{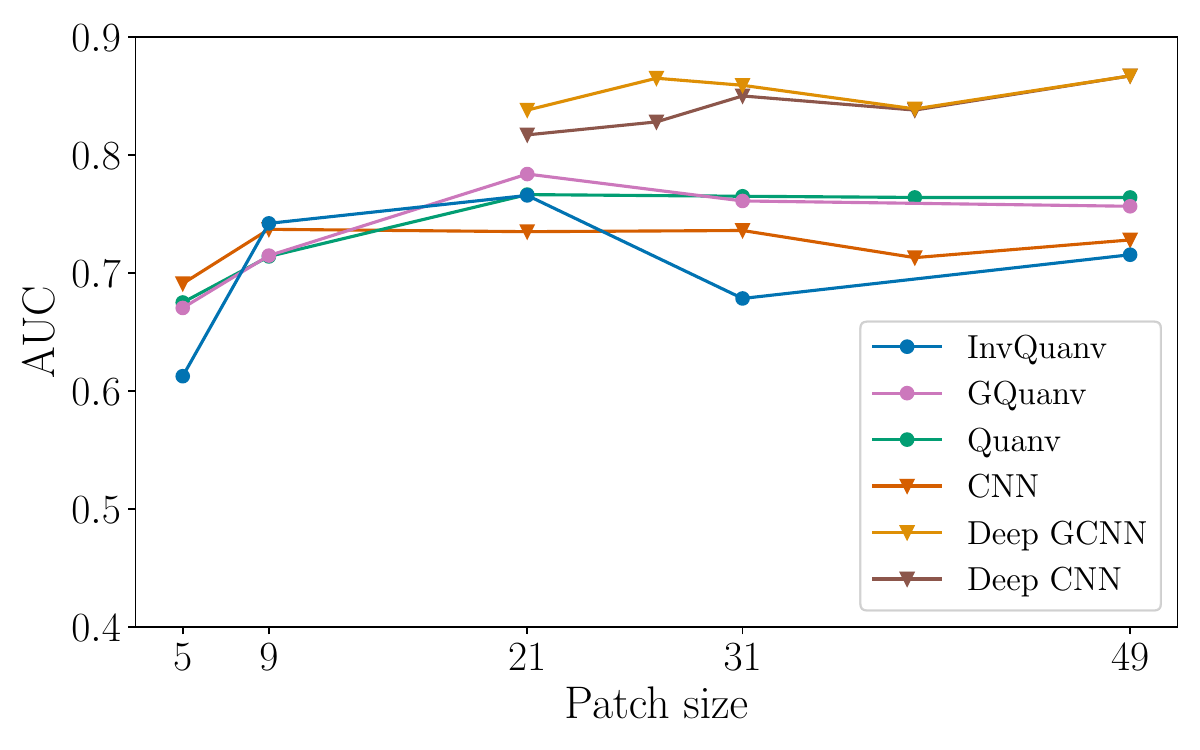}
    \caption{Receiver operating characteristic area under curve (ROC AUC) for all models used in this study with increasing pixel patch size. The deep models were not defined for small patches as no padding was used and it would reduce the size of feature maps to $1 \times 1$ before classification.}
    \label{fig:patch_size_plot_full}
\end{figure}
\leavevmode
\begin{figure}[h!]
    \centering
    \includegraphics[width = \linewidth]{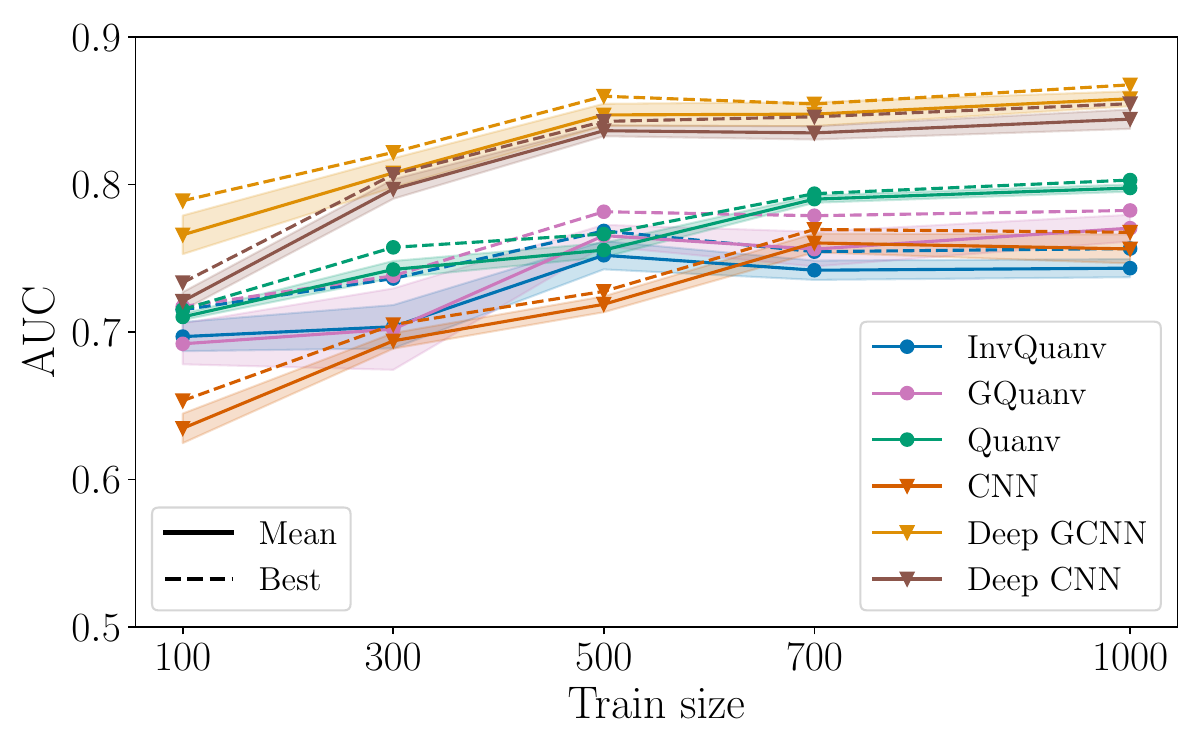}
    \caption{Learning capability of all models used in this study. Vertical axis is the receiver operating characteristic area under curve (ROC AUC) and the horizontal axis is a training size (number of patches). Patch size was kept at $21$.}
    \label{fig:train_size_plot_full}
\end{figure}
\leavevmode
\begin{figure}[h!]
    \centering
    \includegraphics[width = \linewidth]{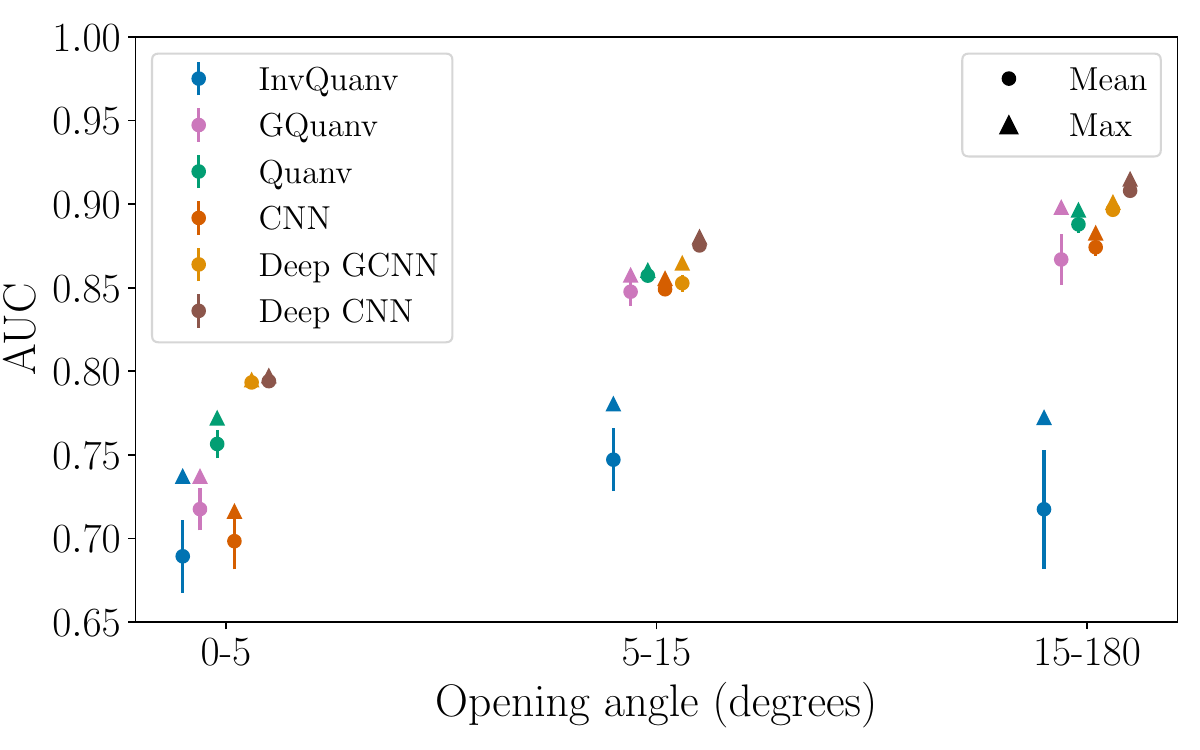}
    \caption{Performance in different environments of all models used in this study. Vertical axis is the receiver operating characteristic area under curve (ROC AUC) and the horizontal axis is the opening angle between the electron and muon particles. The patch size was kept at $21$.}
    \label{fig:particle_bomb_plot_full}
\end{figure}

\newpage
\section{The misinterpreted quanvolution}
\label{sec:Appendix_mis_quanv}
As hinted at in Section \ref{sec:models}, the quanvolution layer used in this study does not - contrary to common interpretation - implement a quantum circuit as a ``filter" in an otherwise classical convolution.

Recalling Equation \ref{eq:convolution}, we note that the convolution involves a dot product of two functions defined on a common space of group elements, where one of the functions is ``moved" through that space.

In order for the quantum circuit to act as the filter, it would have to be defined on the pixels of the feature map \cite{cohen2019general}. One should be able to ask ``What is the value of the quantum circuit at pixel $(2,2)$?".  From the construction of the quanvolution defined in the paper it is clear that such question is meaningless without first finding the value of the feature map at that point - the ``value of the circuit" is necessarily defined through the feature map which becomes embedded into it.

One \textit{can} think of the quantum circuit as a filter in the image-processing sense, a predefined operation applied to a neighbourhood of a pixel \cite{gonzalez_woods}. This should not, however, be extended to thinking of the circuit as the filter of a convolution, based on the above argument.

\subsection{Where is the convolution?}

As reasoned above, the circuit itself cannot be thought of as a filter in a convolution. The quanvolution, however \textit{feels} like a convolution; it seems to inherit its equivariance property, has a stride, a window size and can be implemented, algorithmically, in a similar way to a convolution. It also, as shown in the main text, seems to very naturally extend to group-quanvolutions based on group-convolutions. One way of understanding the connection between the two is to consider how the feature map of a layer is embedded into the quantum circuit. 

\begin{figure}
    \centering
    \includegraphics[trim= 6cm 1cm 6cm 2cm, width=0.5\linewidth]{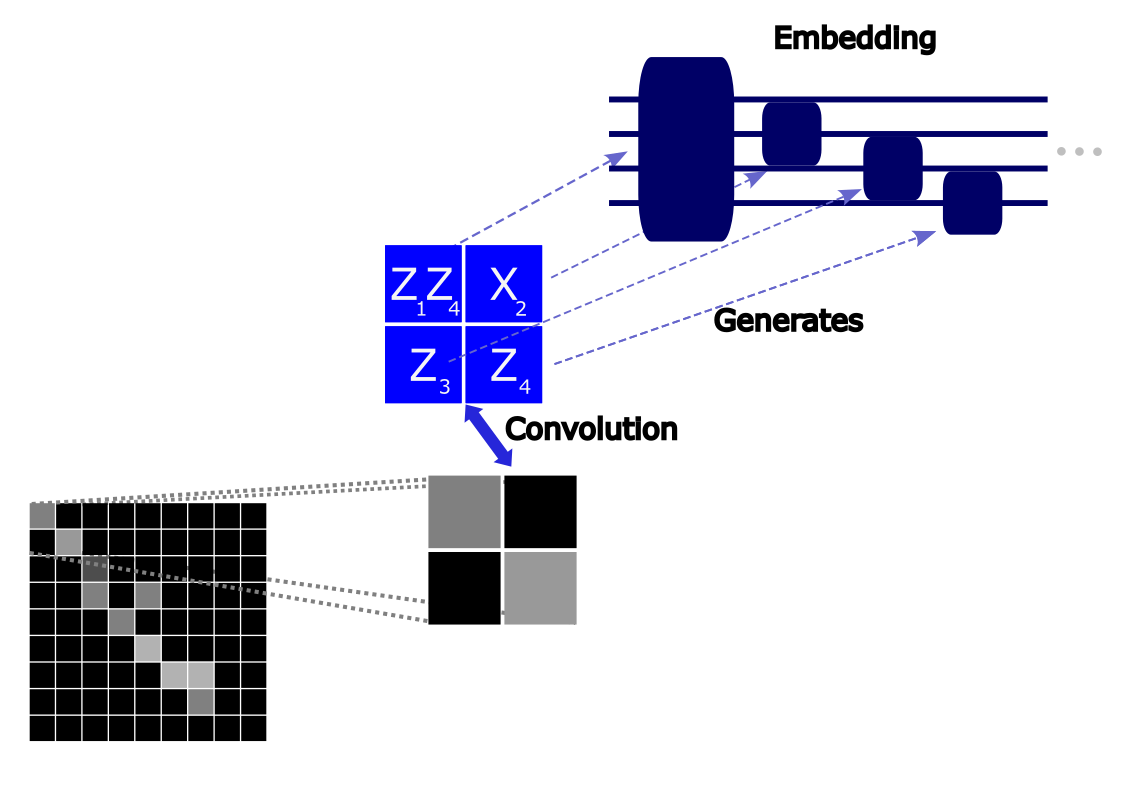}
    \caption{The data-embedding unitary pictured as a convolution with a Hermitian-matrix-valued filter.}
    \label{fig:embedding_with_convolution}
\end{figure}

Let us consider a convolution filter $W$ which is \textit{matrix-valued}. Observe now that if these matrices $W(g)$ are restricted to being Hermitian and commuting with each other, the embedding can be expressed as:
\begin{equation}
\label{eq:embedding_with_convolution}
\begin{split}
    U_{F,W}(g) &= \exp{\left(-i\sum_{h}F(h)W(g^{-1}h)\right)} \\
    &=\exp\left(-i\left[F \star W\right](g)\right).
\end{split}
\end{equation}
The embedding circuit is \textit{generated} by a convolution between the feature map $F$ and a ``quantum filter" $W$. Figure \ref{fig:embedding_with_convolution} demonstrates this idea.
The simple rotation embedding used in the main text (Equation ~\ref{eq:rot_embedding}) can be achieved with a filter containing one of the $\{X_i \ | \ i \in [n_Q]\}$ gates at each entry.
\leavevmode

With Equation~\ref{eq:embedding_with_convolution}, the quanvolution operation can be written as:
\begin{equation}
\begin{split}
\left[F \star^{Q} W\right](g) &= \text{Circ}(U_{F,W}(g)) \\
&=\text{Circ}\left(\exp(-i\left[F \star W\right](g)\right),
\end{split}
\end{equation}
where the $\text{Circ}(U(g))$ function obtains a real-valued property of a state which depends on $g$ only through $U_{F,W}(g)$.
An example would be a reuploader circuit utilising expectation values, like the one used in the main text:
\begin{equation}
\begin{split}
    &\text{Circ}(U_{F,W}(g)) =\\
    &=\langle \psi_0|(U_\theta^{\dagger}U_{F,W}^{\dagger}(g))^{R}|M|(U_{\theta}U_{F,W}(g))^{R}|\psi_0 \rangle.
    \end{split}
\end{equation}

An example of a circuit which does not satisfy the above constraint would be:
\begin{equation}
\begin{split}
    &\text{Circ}(U_{F,W}(g)) =\\
    &= \langle \psi_0|(U_\theta^{\dagger}U_{F,W}^{\dagger}(g))^{R}|M_g|(U_{\theta}U_{F,W}(g))^{R}|\psi_0 \rangle,
    \end{split}
\end{equation}
where the observable $M$ depends on the position in the produced feature map.

To illustrate how the quanvolution inherits the equivariance property from a convolution without being a convolution itself, let us begin with an important property of feature maps in $G-$convolutions:
\begin{equation}
\label{eq:feature_map_property}
    gF(x) = F(g^{-1}x).
\end{equation}
This simply states that to access a value of the $g$-transformed feature map one can look up that value in the original feature map at the point which transforms to $x$ via $g$.
Now, we show a well-known proof for the equivariance of a $G$-convolution \cite{cohen2016group, cohen2019general}:

\begin{equation}
\label{eq:convolution_equivariance}
\begin{split}
    \left[uF \star W\right](g) &= \sum_{h\in G}F(u^{-1}h)W(g^{-1}h)\\
   &= \sum_{h\in G}F(h)W(g^{-1}uh), \quad (h \rightarrow uh) \\
   &= \sum_{h\in G}F(h)W((u^{-1}g)^{-1}h) \\
   &=  \left[F \star W\right](u^{-1}g)\\
   &=  u\left[F \star W\right](g),
\end{split}
\end{equation}

where the last line holds as the convolution defines a new feature map.

Now for a quanvolution:
\begin{equation}
\begin{split}
\left[uF \star^{Q} W\right](g) &= \text{Circ}(U_{uF,W}(g)) \\
& =\text{Circ}(\exp(-i\left[uF \star W\right](g))) \\
& =\text{Circ}(\exp(-i\left[F\star W\right](u^{-1}g))) \\
 &=\left[F \star^{Q} W\right](u^{-1}g) \\
 &=u\left[F \star^{Q} W\right](g),
\end{split}
\end{equation}
where the third line comes from \ref{eq:convolution_equivariance}, and again in the last line, we used the property of feature maps (\ref{eq:feature_map_property}).
This illustrates how any $G$-quanvolution can be obtained from an appropriate $G$-convolution and explains why the proposed $C_4$-quanvolution works.

\section{Details of hyperparameter optimisation}
\label{sec:Appendix_hypopt}

For training and hyperparameter optimisation, \texttt{ray} \cite{moritz2018ray} and \texttt{weightsandbiases} (\texttt{wandb}) \cite{wandb} packages were used extensively.

In the patch size study, for each of the patch sizes, a hyperparameter search over around 200 models per architecture has been performed. Models were allowed to train for up to 100 epochs but were often cut short by the \texttt{RayScheduler} object used in the pipeline. The \texttt{RayScheduler} class contains methods for speeding up wide model searches by cutting jobs short based on a metric of choice. For this study, models were terminated based on the loss they achieved during training.

Loss, validation loss and validation accuracy were reported throughout training to online and local \texttt{wandb} project directories, which helped with visual assessing of their performance and accessing trained models.
After training, the model used for testing was chosen based on the highest validation accuracy achieved during training.
\leavevmode
\begin{table*}[t!]
    \centering
    \caption{Hyperparameters optimised for the models used in the study. $ []$ signifies a continuous range and $\{\}$ a discrete set.}
    \label{tab:hyperparams}
    \begin{tabular}{p{0.37\linewidth}|p{0.22\linewidth}|p{0.3\linewidth}}
    Hyperparameter & Sampled set & Models \\ \hline
    \rule{0pt}{3ex}Learning rate & $[10^{-3}$ , $10^{-1}]$ & All\\
    number of layers (quanvolution circuits) &$\{1,2\}$ & QECNE, QEQE, QNECNE\\
    number of filters (quanvolution) & $\{1,2,3\}$& QECNE, QEQE, QNECNE\\
    number of reuploads (quanvolution circuits) &$\{1,2\}$ & QECNE, QEQE, QNECNE\\
    maximum for parameter initialisation (quanvolution circuits) &$\{10^{-3}, 10^{-1}, \frac{\pi}{4}, \frac{\pi}{2}, 2\pi\}$ & QECNE, QEQE, QNECNE\\
    dense units sizes &$\{(8,8), (128,32)\}$ & QECNE, QNECNE\\
    use of dropout & $\{True, False\}$ & QECNE, QNECNE\\
    dropout amount &$\{10^{-4}, 0.5\}$ & QECNE, QNECNE\\
    classifier number of layers &$\{1,2,3,4\}$ & QEQE\\
    classifier number of reuploads & $\{1,2,3\}$ & QEQE\\
    number of 1-local gates &$\{1,2,3,4\}$  & QEQE\\
    number of 2-local gates &$\{1,2,3,4\}$  & QEQE\\
    placement of 1-local gates & See Appendix \ref{sec:Appendix_geo_class} & QEQE \\
    placement of 2-local gates & See Appendix \ref{sec:Appendix_geo_class} & QEQE \\
    \end{tabular}
\end{table*}


\section{Details of the quantum geometric classifier}
\label{sec:Appendix_geo_class}
The geometric classifier circuits are invariant to rotations of the feature maps received by them (coming from the last layer of the quanvolution and potential pooling).
These classifier circuits, for ease of software implementation, were defined only on $9$-qubit circuits. That means that only $3\times3$ patches could be classified. This provided a restriction for the feature extraction layers (See Figure \ref{fig:quanvolution}) preceding the quantum classifier. That is, the combination of the quanvolutions and the classical pooling had to result in a final layer that has a shape $3\times3$ (before being flattened).

The invariant classifiers' layers have been  obtained according to the following protocol: 
\begin{itemize}
    \item randomly pick a number $N_1$ of single-qubit gates (See Table \ref{tab:hyperparams}) (e.g. $1$)
    \item randomly pick a number $N_2$ of two-qubit gates (See Table \ref{tab:hyperparams})  (e.g. $1$)
    \item randomly pick a qubit for each of the 1-qubit gates to act on (e.g. qubit $3$)
    \item randomly pick a pair of qubits for each of the 2-qubit gates to act on (e.g. qubits $3$ and $6$).
    \item for each of the gates, randomly choose a Pauli string of the correct length (e.g. $X$, $XY$), creating generators ($X_3$, $X_3Y_6$)
    \item by twirling the generators, obtain generators of $C_4$-equivariant gates ($X_1+X_3+X_5+X_7$, $X_3Y_6+X_1Y_0+X_5Y_2+X_7Y_8$)
    \item reject the choice if the resulting generators cannot be implemented as combinations of $2$-local gates, accept otherwise
\end{itemize}
\leavevmode
\begin{figure}
    \centering
    \includegraphics[trim= 6cm 3cm 6cm 3cm,width=0.5\linewidth]{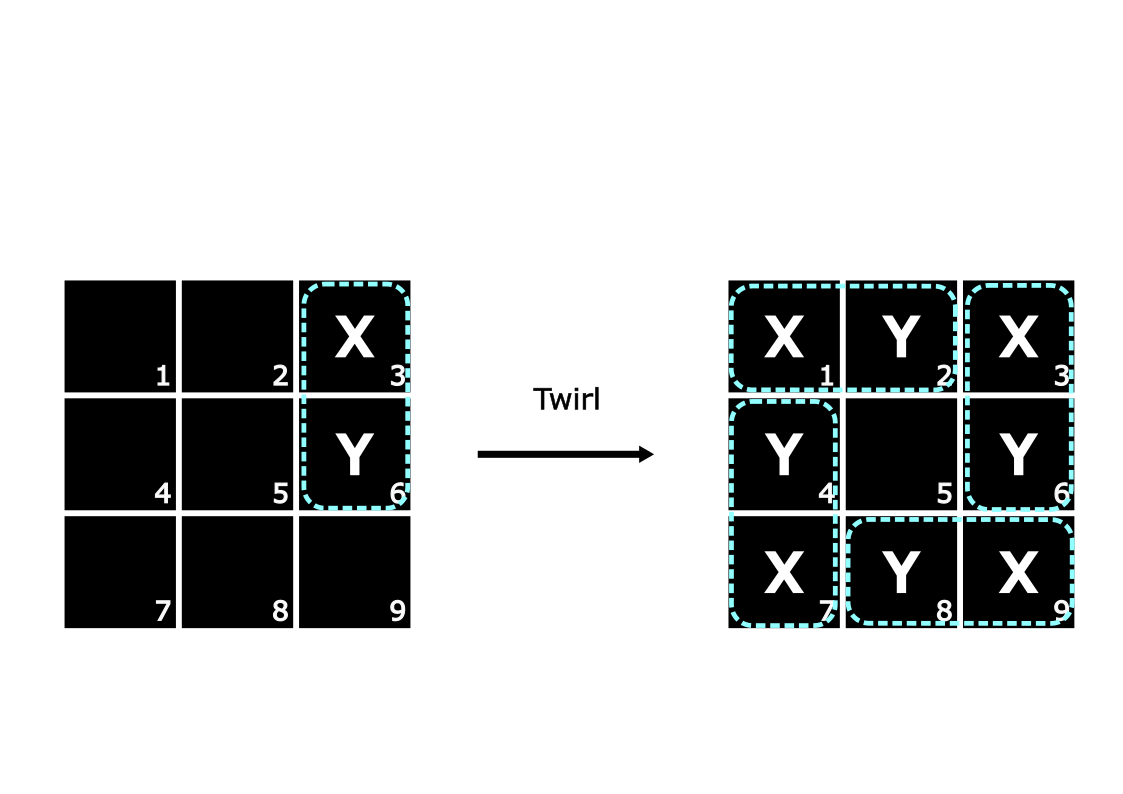}
    \caption{A $C_4$-invariant gate obtained from twirling $X_3Y_6$ overlaid on a pixelised grid representing the input layer to the classifier defined in the text.}
    \label{fig:twirl}
\end{figure}
The rejection criterion in the last step refers to a situation like:
\begin{equation}
    \tau(X_0Z_2) = X_0Z_2 + X_2Z_8 + X_8Z_6 + X_6Z_0.
\end{equation}
Because the elements of this generator do not commute, the resulting parametrised gate cannot be implemented as a product of $2$-qubit gates, $e^{i\theta\tau(X_0Z_2)} \neq e^{i\theta X_0Z_2}e^{i\theta X_2Z_8}e^{i\theta X_8Z_6}e^{i\theta X_6Z_0}$.

We note a detailed study of rotation-invariant circuits for images in~\cite{Sein_2025}, where the feature extraction is achieved using equivariant convolutions (our work uses equivariant quanvolutions).
\bibliographystyle{apsrev4-2}
%

\end{document}